\documentclass[journal,10pt]{IEEEtran}
\usepackage{amsmath}
\usepackage{amssymb}
\usepackage{graphicx}
\usepackage{booktabs}% http://ctan.org/pkg/booktabs
\newcommand{\tabitem}{~~\llap{\textbullet}~~}
\usepackage{multirow}

\makeatletter
\usepackage{amsthm}
\usepackage{epstopdf}
\usepackage{epsfig}
\usepackage{subfig}
\usepackage{cite}\usepackage{calc}%%\linespread{1.6}
\usepackage{cite}%%\newtheorem{theorem}{Theorem}[section]

\usepackage{cite}
\usepackage{hyperref}
\hypersetup{breaklinks=true}

\renewcommand{\figurename}{Fig.}
    \usepackage[font=footnotesize,labelsep=period]{caption}
\makeatother

\begin{document}

\title{Uplink HARQ for Cloud RAN via Separation of Control and Data Planes}

\author{Shahrouz Khalili, \emph{Student Member, IEEE} and\thanks{This work was partially supported by the U.S. NSF grant CCF-$1525629$. Part of the material was presented in \cite{itashah}.}
\thanks{Copyright (c) 2015 IEEE. Personal use of this material is permitted. However, permission to use this material for any other purposes must be obtained from the IEEE by sending a request to pubs-permissions@ieee.org.}
\thanks{S. Khalili and O. Simeone are with CWCSPR, ECE Dept, NJIT, Newark,
USA. E-mail: \{sk669, osvaldo.simeone\}@njit.edu.%
}Osvaldo Simeone, \emph{Fellow, IEEE}}

\maketitle
\begin{abstract}
The implementation of uplink HARQ in a Cloud- Radio Access Network RAN (C-RAN) architecture is constrained by the two-way latency on the fronthaul links connecting the
Remote Radio Heads (RRHs) with the Baseband Units (BBUs) that perform
decoding. To overcome this limitation, this work considers an architecture based on the \textit{separation of control and data} planes, in which retransmission control decisions are made at the edge of the network, that is, by the RRHs or User Equipments (UEs), while data decoding is carried out remotely at the BBUs. This solution enables \textit{low-latency local retransmission decisions} to be made at the RRHs or UEs, which are not subject to the fronthaul latency constraints, while at the same time leveraging the decoding capability of the BBUs.

A system with BBU Hoteling system is considered first in which each RRH has a dedicated BBU in the cloud. For this system, the control-data separation leverages \textit{low-latency local feedback} from an RRH to drive the HARQ process of a given UE. Throughput and probability of error of this solution are analyzed for the three standard HARQ modes of Type-I, Chase Combining and Incremental Redundancy over a general fading MIMO link. Then, novel \textit{user-centric low-latency feedback} strategies are proposed and analyzed for the C-RAN architecture, with a single centralized BBU, based on limited ``hard'' or ``soft'' local feedback from the RRHs to the UE and on retransmission decisions taken at the UE. The analysis presented in this work allows the optimization of the considered schemes, as well as the investigation of the impact of system parameters such as HARQ protocol type, blocklength and number of antennas on the performance of low-latency local HARQ decisions in BBU Hoteling and C-RAN architectures.
\end{abstract}
% For peer review papers, you can put extra information on the cover% page as needed:% \ifCLASSOPTIONpeerreview% \begin{center} \bfseries EDICS Category: 3-BBND \end{center}% \fi% For peerreview papers, this IEEEtran command inserts a page break and% creates the second title. It will be ignored for other modes.%\section{System Model}

\section{Introduction}
\label{sec:intro}
\begin{figure}
\centering \includegraphics[scale=0.26]{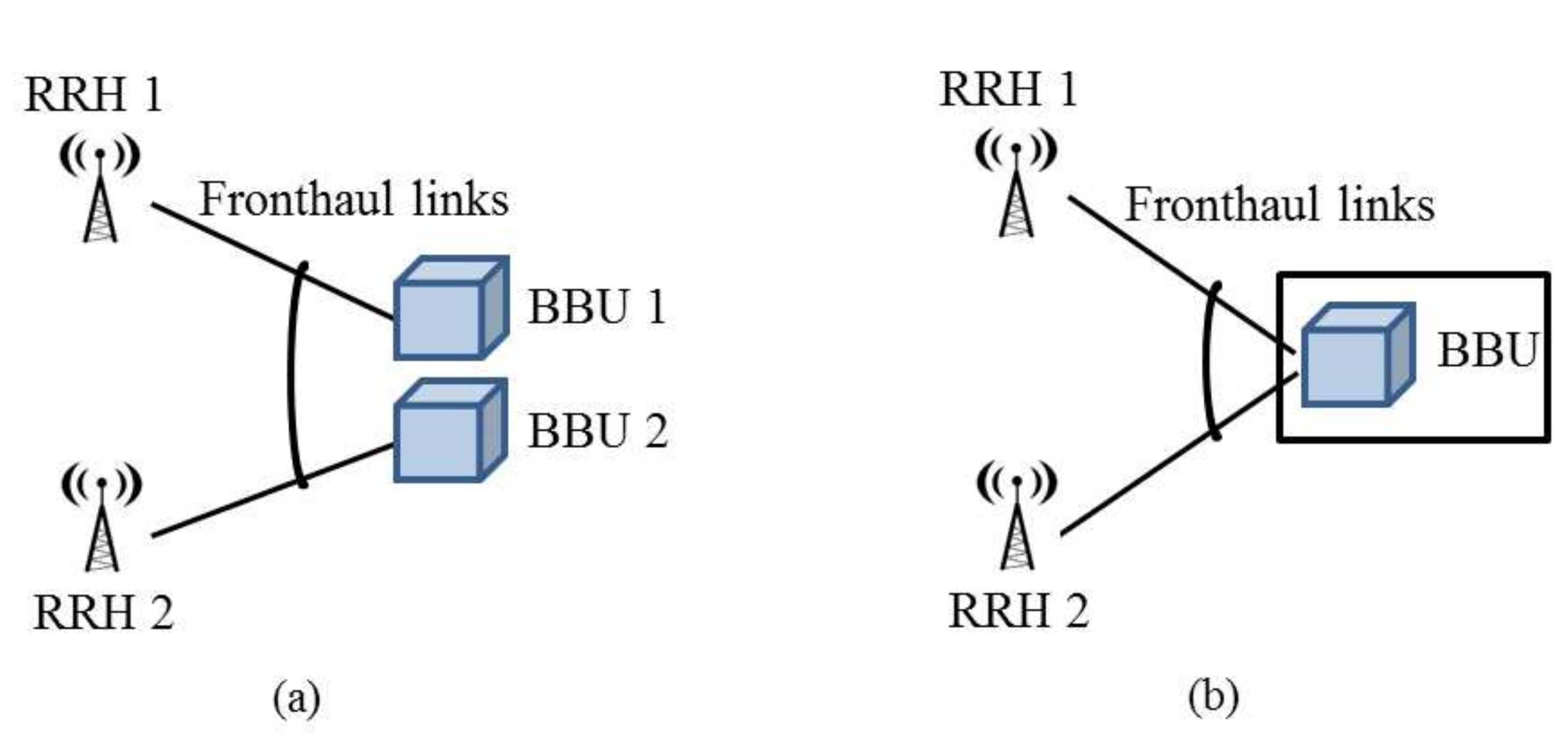} \protect\caption{Illustration of the (a) BBU Hoteling and (b) C-RAN architecture ($L=2$ RRHs).}
\label{dcran}
\end{figure}

The Cloud-Radio Access Network (C-RAN) is a candidate cellular architecture for 5G systems, which is characterized by the separation of each base station into a Remote Radio Head (RRH) that retains only radio functionalities and a Baseband Unit (BBU) that implements the rest of the protocol stack, including the physical layer. In a C-RAN as seen in Fig. \ref{dcran}-(b), a unique BBU is shared among multiple RRHs, enabling joint baseband processing to be carried out at the BBU across the baseband signals of all the connected RRHs. The term ``BBU Hoteling'' is instead used here to refer to an intermediate architecture, shown in Fig. \ref{dcran}-(a), in which the BBUs of different RRHs are distinct, with each BBU performing baseband processing for one RRH (see \cite{hotel,c-ran,check,nahas}). In both cases, the connection between an RRH and a BBU is known as fronthaul link.

The BBU Hoteling and C-RAN architectures lower the expenditure needed to deploy and operate dense cellular networks by simplifying the base stations hardware and by enabling flexible upgrading and easier maintenance (see, e.g., \cite{c-ran,check,nahas}). BBU Hoteling allows limited forms of cooperation to be implemented among base stations in case the BBUs are co-located,  particularly in the downlink, by leveraging an X2 interface that may connect the BBUs with one another within the same cloud \cite{nahas}. Nevertheless, joint baseband decoding in the uplink is generally not feasible with BBU Hoteling, since it requires the exchange of baseband signals among BBUs, rather than user-plane data as allowed by an X2 interface (see e.g., \cite{c-ran,check,nahas}). In contrast, the C-RAN architecture can also benefit from the statistical multiplexing and interference management capabilities that are made possible by joint baseband processing.

\begin{figure}
\centering \includegraphics[scale=0.35]{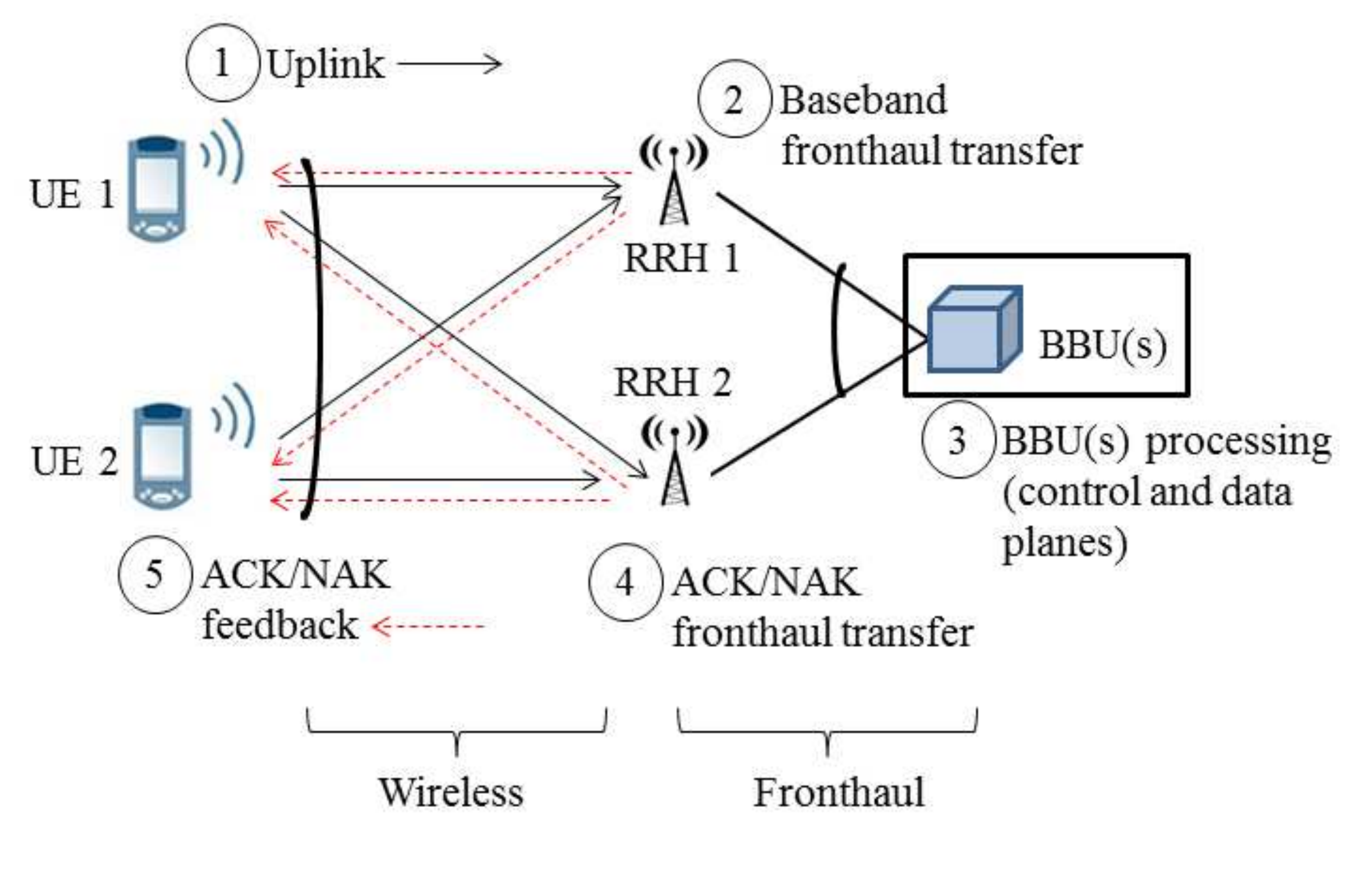} \protect\caption{Conventional HARQ in BBU Hoteling or C-RAN systems. The numbers indicate the sequence of events associated with a transmission. Fronthaul latency is associated with the fronthaul transmissions at steps 2 and 4 and with the part of BBU processing at step 3 needed to encode and decode transmissions on the fronthaul links. The cross-links in the uplink carry interference in a BBU Hoteling system and useful signals in a C-RAN. The dashed cross-links in the ACK/NAK feedback path are used only in the C-RAN architecture.}
\label{dcran2}
\end{figure}

\textbf{Main Problem:} The implementation of the BBU Hoteling and C-RAN architectures needs to contend with the potentially significant latencies needed for the transfer and processing of the baseband signals on the fronthaul links to and from the BBU(s) \cite{ngm}. The communication protocols that are most directly affected
by fronthaul delays are the Automatic Repeat Request (ARQ) and Hybrid ARQ (HARQ)\footnote{In ARQ protocols, different transmissions of a packet are performed independently, whereas, in HARQ schemes, decoding and/or coding can be performed across multiple retransmissions \cite{arq}.} protocols at layer 2. In fact, in a conventional cellular network,
upon receiving a codeword from an UE, the
local base station performs decoding, and, depending on the decoding
outcome, feeds back an Acknowledgment (ACK) or a
Negative Acknowledgment (NAK) to the UE. In contrast, with BBU Hoteling or C-RAN, as illustrated in Fig. \ref{dcran2},
the outcome of decoding at the BBU may only become available at the
RRHs after the time required for the transfer of the
baseband signals from the RRHs to the BBU(s) on the fronthaul links, for processing at the BBU(s), and
for the transmission of the decoding outcome from the BBU(s) to the RRHs on the fronthaul links.

The fronthaul latency may significantly affect the performance of retransmission protocols. For instance, in LTE with frequency division
multiplexing, the feedback latency should be less than $3$ ms in
order not to disrupt the operation of the system \cite{rrh-bell}\footnote{By interleaving multiple HARQ processes, as discussed in \cite{rrh-bell}, the tolerated latency can be increased to $3+n8$ ms, where $n$ is a positive integer, albeit at the cost of possibly reducing the throughput. Of the mentioned $3$ ms latency, it has been recently specified that the one-way transport delay on the fronthaul should be no larger than around $400$ $\mu$s \cite{ngm}.}. We also refer to \cite{gulati,osiarc} for a discussion on the effect of the latency on HARQ and ARQ protocols in C-RANs.

\textbf{A Solution Based on the Separation of Control and Data Planes:} Fronthaul latency is unavoidable in conventional BBU Hoteling and C-RAN architectures in which the RRHs only retain radio functionalities. Nevertheless, alternative functional splits are currently being investigated whereby the RRH may implement some additional functions \cite{rrh-bell,check,rost1,olive}. In this work, we consider a functional split that enables the \emph{separation of control and data planes} associated with the HARQ protocol, with the aim of alleviating the problem of fronthaul latency. We note that the approach studied here can be seen as an instance of the more general principle of control and data separation, for which an overview of the literature can be found in \cite{moham}.

In particular, we investigate an architecture in which retransmission \textit{control} decisions are made at the edge of the network, that is, by the RRHs or UEs, while decoding of \textit{data-plane} information is carried out remotely at the BBUs as in conventional BBU Hoteling or C-RAN systems. This separation of HARQ control at the edge and data-plane processing at the BBU(s) has the following advantages: (\emph{i}) retransmission control is not subject to the fronthaul latency constraints; (\emph{ii}) given that data-plane processing is performed at the BBU(s), the complexity of the RRHs can be kept below that of a conventional base station\footnote{While the computational complexity of the operations carried out at the BBUs is not a primary concern for the BBU Hoteling or C-RAN architectures, it is noted that the solutions proposed here based on the separation of control and data planes, do not increase the complexity of the BBU as compared to that of a conventional system.}; (\emph{iii}) the benefits of joint baseband processing of data-plane information at the BBU of a C-RAN system in terms of spectral efficiency are maintained.

The implementation of low-latency local control of the retransmission process at the edge is made possible by an RRH-BBU functional split whereby each RRH can perform synchronization and resource demapping \cite{rrh-bell}\footnote{In an OFDM system, such as LTE, this requires also the implementation of an FFT block.}, so as to perform the estimation of uplink channel state information (CSI). To elaborate, consider first a BBU Hoteling system. As proposed in  \cite{rrh-bell} and in \cite{rost}, based on the available CSI, an RRH can attempt to predict whether successful or unsuccessful decoding is expected to occur at the BBU for the symbol received from a given UE. Accordingly, it can feed back an ACK/NAK message without waiting to be notified about the actual decoding outcome at the BBU and without running the channel decoder on the data-plane information, which is implemented only at the BBU. Fig. \ref{local-feedback} presents an illustration of the outlined low-latency approach.

\begin{figure}
\centering \includegraphics[scale=0.38]{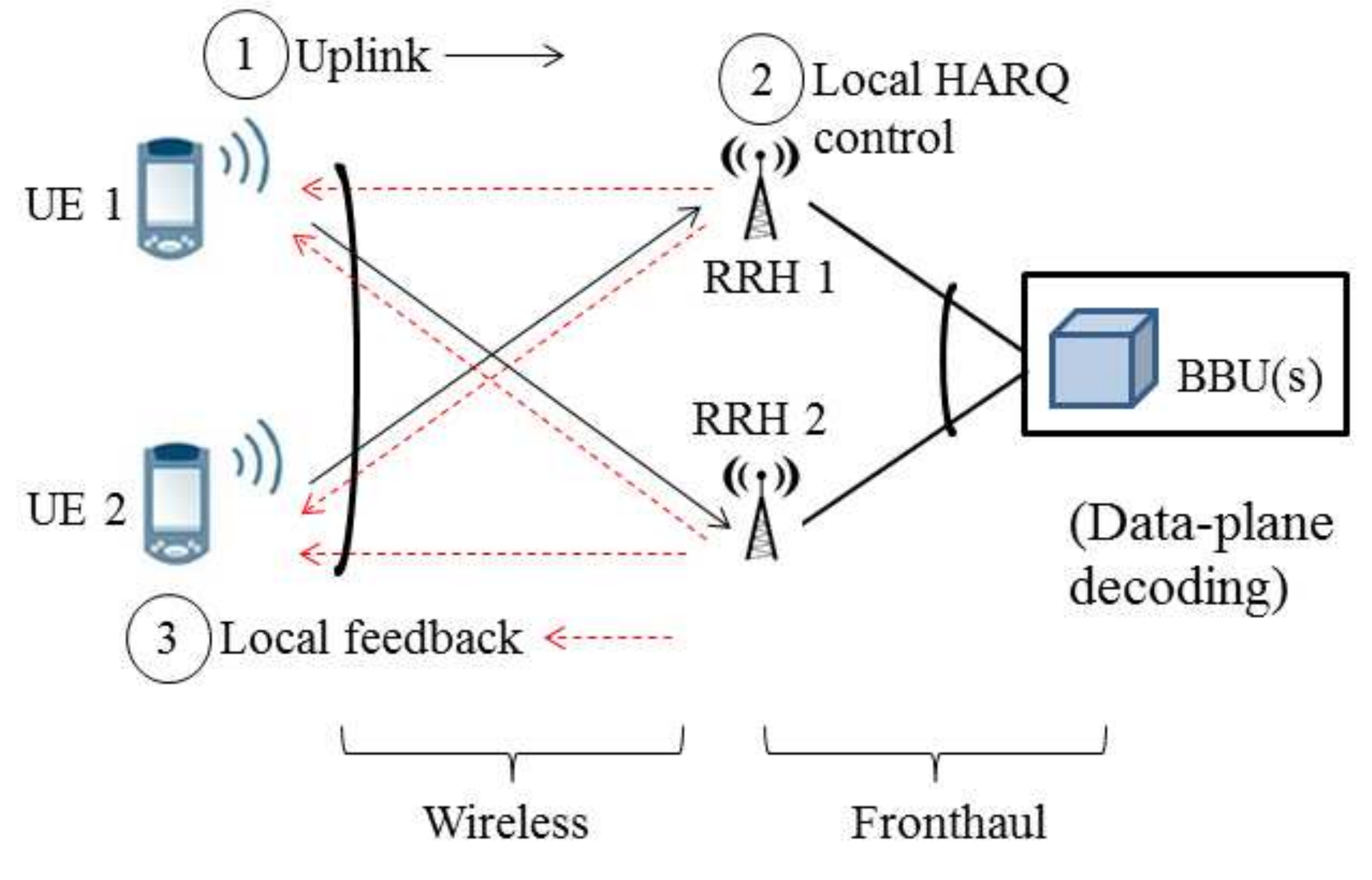} \protect\caption{HARQ in BBU Hoteling and C-RAN systems via low-latency local feedback based on separation of control and data planes. HARQ control is carried out at the network edge based on local low-latency feedback from the RRHs, while data decoding is carried out at the BBUs. The cross-links in the uplink carry interference in a BBU Hoteling and useful signals in a C-RAN. The cross-links in the feedback path are used only in the C-RAN architecture.}
\label{dcran3}
\end{figure}

%\begin{figure}
%\centering \includegraphics[scale=0.40]{rrhbbu} \protect\caption{Uplink HARQ for C-RAN: The switch is closed in case of conventional BBU-based
%feedback while it is open with RRH-based local feedback \cite{rrh-bell}\cite{rost}.}
%\label{rrhbbu}
%\end{figure}
The local feedback approach under discussion introduces
possible errors due to the mismatch between the local decision at
the RRH and the actual decoding outcome at the BBU. Indeed, the RRH
may request an additional retransmissions for a packet that the BBU
is able to decode, or acknowledge correct reception of a packet for
which decoding eventually fails at the BBU, hence causing a throughput
degradation.

In a C-RAN, which is characterized by joint baseband processing across multiple RRHs, the outlined approach based on control at the edge is complicated by the fact that the CSI between the UE and each RRH is not known to other RRHs. Therefore, it is not possible for the RRHs to directly agree on HARQ control decisions, making the local feedback mechanism proposed in \cite{rrh-bell} and \cite{rost} not applicable.

\textbf{Main contributions:} The main contributions of this paper are summarized as follows.
\begin{itemize}
\item For BBU Hoteling, we analyze throughput and probability of error of the outlined approach based on control and data separation for the three standard HARQ modes of Type-I (TI), Chase Combining (CC) and Incremental Redundancy (IR) over a multi-antenna, or MIMO, link with coding blocks (packets) of arbitrary finite length. This is done by leveraging recently derived finite-blocklength capacity bounds \cite{poly}. As a result, unlike the existing literature \cite{rrh-bell} and \cite{rost}, the analysis allows the investigation of the impact of system parameters such as HARQ protocol type, blocklength and number of antennas. We note that the analysis in \cite{rost} focuses on the throughput of single-antenna links in a BBU Hoteling with HARQ-IR and is based on an error exponent framework, which is known to be provide an inaccurate evaluation of the probability of error in the practical finite-blocklength regime \cite[Eq. (54)]{poly}\cite[Sec. 1.2 and Sec. 1.3]{yuri}.
\item We propose and analyze \emph{user-centric low-latency feedback} schemes for C-RAN systems based on the control and data separation architecture. According to the proposed techniques, limited-feedback information is sent from each RRH to an UE in order to allow the latter to make a low-latency local control decision about the need for a retransmission.  A ``hard feedback'' approach is first proposed that directly generalizes the BBU Hoteling scheme described above and requires a one-bit feedback message from each RRH. Then, a ``soft feedback'' strategy is proposed in which the UE decision is based on multi-bit feedback from the RRHs, consisting of quantized local CSI.
\end{itemize}
The rest of the paper is organized as follows. In Sec. \ref{sec:sys}, the system model for BBU Hoteling and C-RAN systems is introduced. Sec. \ref{ran} details the principles underlying the proposed low-latency local feedback solutions for BBU Hoteling and C-RAN systems. The metrics used to evaluate the performance of the proposed schemes and some preliminaries are discussed in Sec. \ref{per-cert}. In Sec. \ref{dran:an} and Sec. \ref{cran:an}, the analysis of BBU Hoteling and C-RAN strategies is presented. In Sec. \ref{sec:sim}, the numerical results are provided, and Sec. \ref{sec:con} concludes the paper.

\textit{Notation}: Bold letters denote matrices and superscript $^H$ denote Hermitian conjugation. $\mathcal{CN}(\mu,\sigma^2)$ denotes a complex normal distribution with mean $\mu$ and variance $\sigma^2$; and $\mathcal{X}^2_k$ a Chi-Squared distribution with $k$ degrees of freedom. $f_{\mathcal{A}}(x)$ and $F_{\mathcal{A}}(x)$ represent the probability density function and the cumulative distribution function of a distribution $\mathcal{A}$ evaluated at $x$, respectively. $\textbf{A}=\mathrm{diag}([\textbf{A}_1,...,\textbf{A}_n])$ is a block diagonal matrix with block diagonal given by the matrices $[\textbf{A}_1,...,\textbf{A}_n]$. The indicator function $\textbf{1}(x)$ equals $1$ if $x=\mathrm{true}$ and $0$ if $x=\mathrm{false}$.
\section{System Model}
\label{sec:sys}

We study the uplink of both BBU Hoteling and C-RAN systems as illustrated in Fig. \ref{dcran}. In this section, we detail system model and performance metrics.

\subsection{System Model}
As seen in Fig. \ref{dcran}, each RRH is connected by means of orthogonal fronthaul links to a dedicated BBU for BBU Hoteling and to a single BBU for C-RAN systems. The BBUs perform decoding, while the RRHs have limited baseband processing functionalities
that allow resource demapping and the inference of CSI as discussed in Sec. \ref{sec:intro} and further detailed below. Different UEs are served in distinct time-frequency resources, as done for instance in LTE, and hence we focus here on the performance of a given UE.

Each packet transmitted by the UE contains $k$ encoded complex symbols and is transmitted within a
coherence time/frequency interval of the channel, which is referred
to as \emph{slot}. The transmission rate of the first transmission
of an information message is defined as $r$ bits per symbol, or equivalently, bit/s/Hz, so that $kr$ is the number of
information bits in the information message.

Each transmitted packet is acknowledged via the transmission of a feedback message by the
RRHs. We assume that these feedback
messages are correctly decoded by the UE. We will first assume that messages are limited to binary positive or negative acknowledgments, i.e., ACK or NAK messages, in Sec. \ref{dran}, and we will consider the more general case in which feedback messages may consist of $b\geq1$ bits in Sec. \ref{cran}. The same information message may be transmitted for up to $n_{max}$ successive
slots using standard HARQ protocols such as TI, CC and IR, to be recalled
in Sec. \ref{ran}.

The UE is equipped with $m_{t}$ transmitting antennas, while $m_{r,l}$ receiving antennas are available at the $l$th RRH. The received
signal for any $n$th slot at the $l$th RRH can be expressed as
\begin{equation}\label{fist-sig}
\textbf{y}_{l,n}=\sqrt{\frac{s}{m_{t}}}\textbf{H}_{l,n}\textbf{x}_n+\textbf{w}_{l,n},
\end{equation}
where $s$ measures the average SNR per receive antenna; $\textbf{x}_n\in\mathbb{C}^{m_{t}\times1}$
represents the symbols sent by the transmit antennas at a given channel
use, whose average power is normalized as $\mathrm{E}[||\textbf{x}_n||^{2}]=1$;
$\textbf{H}_{l,n}\in\mathbb{C}^{m_{r,l}\times m_{t}}$ is the channel matrix,
which is assumed to have independent identically distributed (i.i.d.)
$\mathcal{CN}(0,1)$ entries (Rayleigh fading); and $\textbf{w}_{l,n}\in\mathbb{C}^{m_{r,l}\times1}$
is an i.i.d. Gaussian noise vector with $\mathcal{CN}(0,1)$ entries.
The channel matrix $\textbf{H}_{l,n}$ are independent for different RRHs $l\in\{1,...,L\}$ and also change
independently in each slot $n$. Moreover, they are assumed to be known to the $l$th RRH and to the BBU. We assume the use of Gaussian codebooks with an equal power allocation across the transmit antennas, although the analysis could be extended to arbitrary power allocation and antenna selection schemes.

\subsection{Performance Metrics}
The main performance metrics of interest are as follows.
\begin{itemize}
\item Throughput $T$: The throughput measures the average rate, in bits per symbol, at which information can be successfully delivered from the UE to the BBU;
\item Probability $\mathrm{P_s}$ of success: The metric $\mathrm{P_s}$ measures the probability of a successful transmission within a given HARQ session, which is the event that, in one of the $n_{max}$ allowed transmission attempts, the information message is decoded successfully at the BBU.
      \end{itemize}
Note that errors in the HARQ sessions can be dealt with by higher layers, as done by the RLC layer in LTE \cite{rrh-bell}, albeit at the cost of large delays. For this reason, in Sec. \ref{sec:sim}, we will pay special attention to the throughput that can be obtained under a given constraint on the probability of success $\mathrm{P_s}$. Typical values for $\mathrm{P_s}$, on which one can base the design of higher layers, are in the order of $0.99$ -- $0.999$ \cite{nahas}\cite{lte}. We elaborate on the evaluation of these metrics in Sec. \ref{per-cert}. We finally observe that the overall latency generally depends on the specific system implementation, and, most notably, on the delays associated with transport and processing on the fronthaul. We will provide further discussion on this point in Sec. \ref{per-cert}.

\section{Low-Latency Local Feedback}
\label{ran}
In this section, we introduce the key working principles underlying low-latency local feedback solutions for BBU Hoteling and C-RAN.
\subsection{RRH-Based Low-Latency Local Feedback for BBU Hoteling}
\label{dran}
\begin{figure}
\centering \includegraphics[scale=0.35]{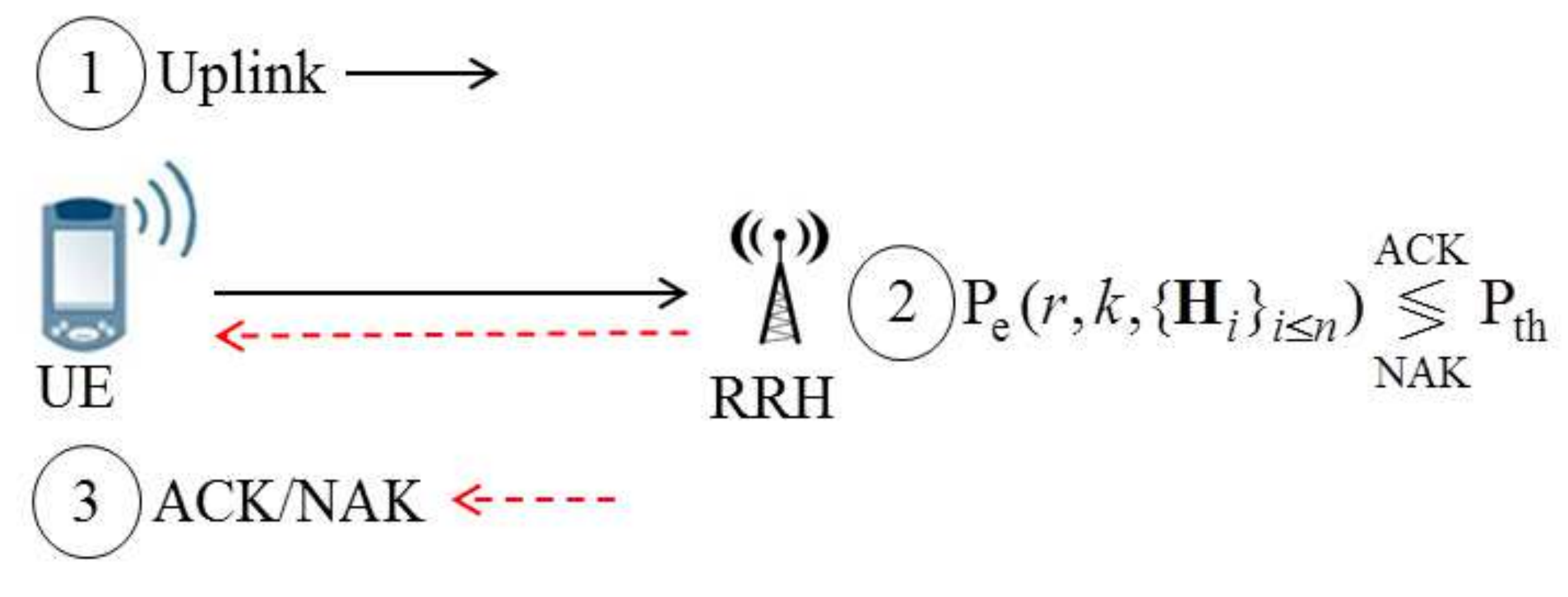} \protect\caption{Low-latency local feedback scheme for BBU Hoteling systems: ACK/NAK messages are sent by the assigned RRH to a given UE according to the local decision rule (\ref{rule_gen}).}
\label{local-feedback}
\end{figure}
In a BBU Hoteling architecture, each pair of RRH and corresponding BBU operates as a base station in a conventional cellular system \cite{c-ran}\cite{check}. Therefore, an UE is assigned to a specific RRH-BBU pair by following standard user association rules. For BBU Hoteling, as in \cite{rost}, we can then focus on a single RRH, i.e., $L=1$, with the understanding that the noise term in (\ref{fist-sig}) may account also for the interference from UEs associated to other RRH-BBU pairs. When studying BBU Hoteling systems, we hence drop the subscript $l$ indicating the RRH index.

The low-latency local feedback scheme for BBU Hoteling, first proposed in \cite{rrh-bell}, is illustrated in Fig. \ref{local-feedback}. At each transmission attempt $n$, the RRH performs resource demapping and obtains CSI about the channel $\textbf{H}_n$. The Modulation and Coding Scheme (MCS) used for data transmission is decided by the BBU during schedualing and can be sent by the BBU to the RRH. Note that the MCS amounts here to the rate $r$ and packet length $k$. Based on this information, the RRH can compute the probability of error $\mathrm{P_{e}}(r,k,\{\textbf{H}_i\}_{i\leq n})$ for decoding at the BBU, where we emphasized the possible dependence of the probability of error $\mathrm{P_{e}}$ on all channel matrices $[\textbf{H}_1,\cdots,\textbf{H}_n]$ corresponding to prior and current transmission attempts. We note that the probability $\mathrm{P_{e}}$ may be read on a look-up table or obtained from some analytical approximations as discussed in the next section. As proposed in \cite{rost}, if the decoding error probability $\mathrm{P_{e}}(r,k,\{\textbf{H}_i\}_{i\leq n})$ is smaller than a given threshold $\mathrm{P}_{\mathrm{th}}$, the RRH sends an ACK message to the UE, predicting a positive decoding event at the BBU; while, otherwise, a NAK message is transmitted, that is,
\begin{equation}\label{rule_gen}
\mathrm{P_{e}}(r,k,\{\textbf{H}_i\}_{i\leq n})\underset{\mathrm{NAK}}{\overset{\mathrm{ACK}}{\lessgtr}}\mathrm{P_{th}}.
\end{equation}

As we will discuss in Sec. \ref{sec:sim}, the optimization of the threshold $\mathrm{P}_{\mathrm{th}}$ needs to strike a balance between the probability of success $\mathrm{P_s}$, which would call for a smaller $\mathrm{P}_{\mathrm{th}}$ and hence more retransmissions, and the throughput $T$, which may be generally improved by a larger $\mathrm{P}_{\mathrm{th}}$, resulting in the transmission of new information.

%In HARQ-TI, the decoding at the BBU is performed only based on the last retransmitted packet and previous transmissions are neglected. In CC, the the previous transmissions are combined to increase the effect SNR to improve the decoding performance. In IR scheme, the UE transmits new parity bits at each transmission attempt and the BBU performs decoding based on all the received packets.

\subsection{User-Centric Low-Latency Local Feedback for C-RAN}
\label{cran}
In C-RAN, unlike BBU Hoteling systems, a BBU jointly processes the signals received by several connected RRHs (Fig. \ref{dcran}-(b)). Therefore, a UE-RRH assignment step is not needed as the BBU performs decoding based on the signals received from all connected RRHs. The development of a local feedback solution based on the implementation of control decision at the edge is hence complicated for C-RAN by the fact that the BBU decoding error probability $\mathrm{P_{e}}(r,k,\{\textbf{H}_i\}_{i\leq n})$ depends on the CSI $\{\textbf{H}_i\}_{i\leq n}$ between the UE and all RRHs, while each RRH $l$ is only aware of the CSI $\{\textbf{H}_{l,i}\}_{i\leq n}$ between the UE and itself. Therefore, the decoding error probability $\mathrm{P_{e}}(r,k,\{\textbf{H}_i\}_{i\leq n})$ cannot be calculated at any RRH as instead done for BBU Hoteling.

\begin{figure}
\centering \includegraphics[scale=0.40]{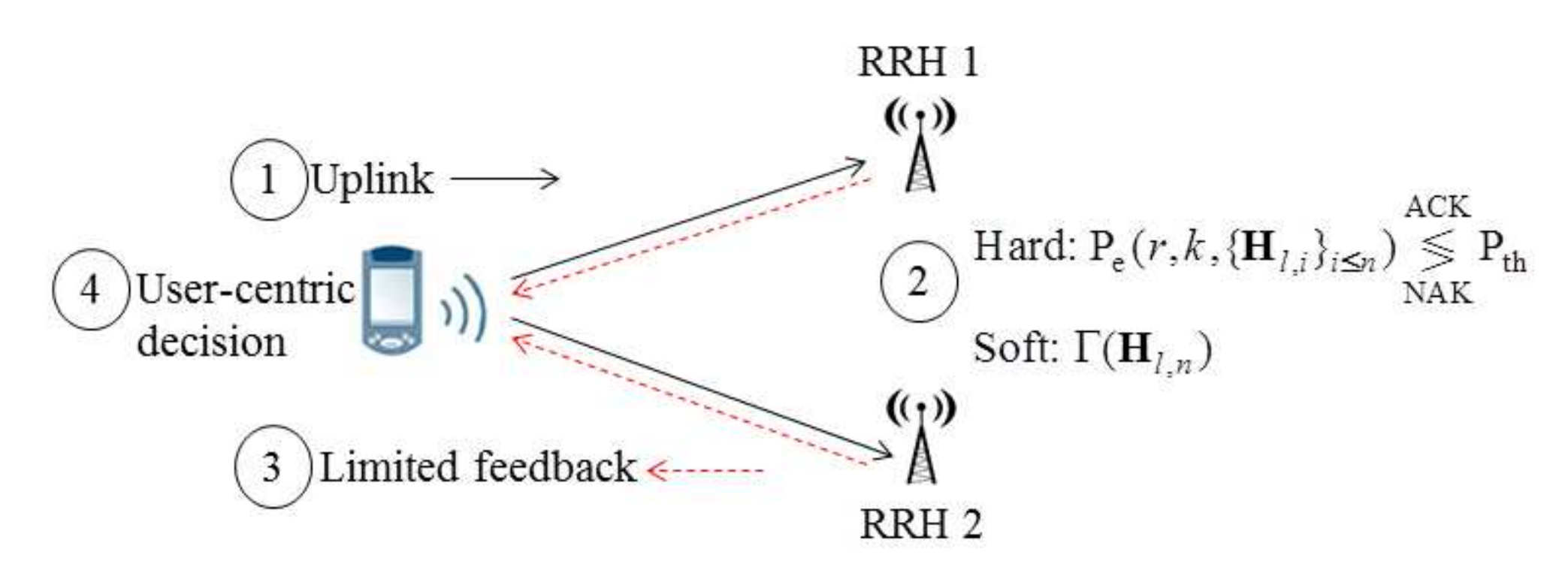} \protect\caption{Low-latency local feedback scheme for C-RAN systems: The UE collects limited-feedback messages from the RRHs to make a local control decision on whether another transmission attempt is necessary.}
\label{cranlocal-feedback}
\end{figure}
To overcome this problem, in this paper, we propose a user-centric low-latency local HARQ mechanism, whereby the UE collects limited-feedback messages from the RRHs, based on which it makes a local decision about whether a further retransmission attempt is needed or not, illustrated in Fig. \ref{cranlocal-feedback}. We allow for multi-bit feedback messages from the RRHs to the UE, and study methods based on $\textit{hard feedback}$, and $\textit{soft feedback}$, as explained next.
\subsubsection{Hard Feedback}
\label{hard:detail}
The hard feedback scheme is a direct extension of the local feedback solution explained in Sec. \ref{dran} for BBU Hoteling. Since at the $n$th transmission attempt, the $l$th RRH is only aware of the CSI $\{\textbf{H}_{l,i}\}_{i\leq n}$ between itself and the UE, it can only calculate the decoding error probability $\mathrm{P_{e}}(r,k,\{\textbf{H}_{l,n}\}_{i\leq n})$, which corresponds to a scenario in which the BBU decodes solely based on the signal received by the $l$th RRH. Then, each RRH $l$ uses a $1$-bit quantizer, which maps the probability $\mathrm{P_{e}}(r,k,\{\textbf{H}_{l,n}\}_{i\leq n})$ to an ACK/NAK message according to the same rule used in BBU Hoteling system, i.e.,
\begin{equation}\label{rule_hard}
\mathrm{P_{e}}(r,k,\{\textbf{H}_{l,n}\}_{i\leq n})\underset{\mathrm{NAK}}{\overset{\mathrm{ACK}}{\lessgtr}}\mathrm{P_{th}}.
\end{equation}
The UE decides to retransmit the packet if all RRHs return a NAK message and to stop retransmissions if at least one ACK is received.
\subsubsection{Soft Feedback}
\label{soft:detail}
The soft feedback schemes aims at leveraging multi-bit feedback messages, composed of $b\geq1$ bits, from each RRH to the UE. The key idea here is that the UE can estimate the decoding error probability $\mathrm{P_{e}}(r,k,\{\textbf{H}_i\}_{i\leq n})$ of the BBU upon receiving information from each RRH $l$ about the local CSI $\textbf{H}_{l,n}$. To this end, in the soft feedback scheme, each RRH quantizes its own CSI $\textbf{H}_{l,n}$ by using vector quantization \cite{quant} with $b$ bits and sends the quantized CSI $\Gamma(\textbf{H}_{l,n})=\hat{\textbf{H}}_{l,n}$ to the UE via a $b$-bit feedback message. Then, the UE performs a retransmission if the estimated decoding error probability $\mathrm{P}_{\mathrm{e}}(r,k,\{\hat{\textbf{H}}_i\}_{i\leq n})$, with $\hat{\textbf{H}}_i$ collecting all the quantized matrices $\hat{\textbf{H}}_{l,n}$ for $l\in\{1,...,L\}$, is larger than a threshold $\mathrm{P_{th}}$ and stop retransmission otherwise, as in
\begin{equation}\label{rule-soft}
\mathrm{P_{e}}(r,k,\{\hat{\textbf{H}}_i\}_{i\leq n})\underset{\mathrm{NAK}}{\overset{\mathrm{ACK}}{\lessgtr}}\mathrm{P_{th}}.
\end{equation}

\section{Performance Criteria and Preliminaries}
\label{per-cert}
In this section, we discuss the general approach that will be followed to evaluate throughput and probability of success for the considered schemes in BBU Hoteling and C-RAN systems. We also discuss the comparison in terms of average latency between the conventional C-RAN implementation and the considered feedback schemes.

\subsection{Throughput and Probability of Success}
To start, let us denote as $\mathrm{RTX}_{n}$ the event that a retransmission decision is made for all the first $n$ transmission attempts of an information
message. In a similar manner, we define as $\mathrm{STOP}_n$ the event that a decision is made to stop the retransmission of a packet at the $n$th attempt, and hence $n-1$ retransmission attempts have been performed before. As we discussed in Sec. \ref{ran}, these decisions are made at the RRH for the low-latency local feedback scheme in BBU Hoteling and at the UE in the proposed user-centric low-latency strategies for C-RAN. By definition, the probabilities of these events satisfy the equality
\begin{equation}\label{p_ack}
\mathrm{P}(\mathrm{STOP}_{n})=\mathrm{P}(\mathrm{RTX}_{n-1})-\mathrm{P}(\mathrm{RTX}_{n}).
\end{equation}

In case of ideal feedback from the BBU, a STOP/RTX event reflects correct/incorrect
decoding at the BBU, whereas this is not the case for the local feedback schemes due to the possible mismatch between the RRHs' or users' decisions and the decoding outcome at the BBU. In particular, there are two types of error as summarized in Table \ref{error-table}. In the first type of error, the transmitted packet is not decodable at the BBU, but a STOP decision is made by the local feedback scheme. This type of mismatch causes a failure of the HARQ process, hence adding to the probability of error, or, equivalently reducing the probability of success. In practice, this event may need to be dealt with by higher layers. In the second type of error, the received packet is decodable at the BBU, but an RTX decision is made. In this case, the UE either performs an unnecessary retransmission, hence increasing the number $N$ of transmissions, or, in case the maximum number $n_{max}$ of retransmissions has already been carried out, the HARQ session fails.

\begin{table}[]
\centering
\caption{Error types due to low-latency local feedback}
\label{error-table}
\begin{tabular}{|c|c|l|}
\hline
\textbf{BBU decoding} & \textbf{Local feedback} & \textbf{Consequence} \\
\textbf{outcome} & \textbf{decision} & \\ \hline
\multirow{2}{4em}{Undecodable} & \multirow{2}{4em}{STOP} &  \tabitem HARQ session failure \\
& & \tabitem Delays due to higher-layer \\
& & ~~~~ protocols\\ \hline
\multirow{2}{4em}{Decodable} & \multirow{2}{4em}{RTX} & \tabitem HARQ retransmission\\
& & \tabitem HARQ session failure if last\\
& & ~~~~transmission\\ \hline
\end{tabular}
\end{table}

We now elaborate on the calculation of the throughput $T$ and probability of success $\mathrm{P_s}$ for both the local feedback schemes and reference ideal case of zero-delay feedback from the BBU. We emphasize that for local feedback, we will consider both mistmach events in Table \ref{error-table}, following the discussion above. For all schemes, based on standard renewal theory arguments,
the throughput can be calculated as \cite{caire}
\begin{equation}\label{thro}
T=\frac{r\mathrm{P_{s}}}{\mathrm{E}[N]},
\end{equation}
where we recall that $r$ is the transmission rate, and the random variable $N$
denotes the number of transmission attempts for a given information message.
The average number of transmissions can be computed directly as
\begin{equation}
\begin{split}\mathrm{E}[N] & =\sum_{n=1}^{n_{max}-1}n\mathrm{P}(\mathrm{STOP}_{n})+n_{max}\mathrm{P}(\mathrm{RTX}_{n_{max}-1}).
\end{split}
\label{ave}
\end{equation}
Moreover, the probability of a successful transmission for the case of zero-delay feedback from the BBU is given as
\begin{equation}
\mathrm{P_{s}}=1-\mathrm{P}(\mathrm{RTX}_{n_{max}}).\label{eq:Ps}
\end{equation}

Instead, with local feedback, a transmission
is considered as successful if a decision is made to stop the retransmission of a packet within one of the
$n_{max}$ allowed transmissions attempts \emph{and} if the
BBU can correctly decode. Hence, by the law of total probability,
the probability of success $\mathrm{P_{s}}$ can be written as
\begin{equation}
\mathrm{P_{s}}=\sum_{n=1}^{n_{max}}\mathrm{P}(\mathrm{D}_{n}|\mathrm{STOP}_{n})\mathrm{P}(\mathrm{STOP}_{n}),\label{ps}
\end{equation}
where $\mathrm{D}_{n}$ is the event that the BBU can correctly decode
at the $n$th transmission.

In summary, in order to evaluate the throughput, we use (\ref{p_ack})-(\ref{ave}) for both ideal and local feedback; while, for the probability of success $\mathrm{P_{s}}$, we use (\ref{eq:Ps}) for the case of ideal feedback and (\ref{ps}) for local feedback. Therefore, to compute both metrics, we only need to calculate the probabilities $\mathrm{P}(\mathrm{RTX}_n)$, for both ideal and local feedback, and the probabilities $\mathrm{P}(\mathrm{D}_{n}|\mathrm{STOP}_{n})$ for local feedback, with $n=1,...,n_{max}$. We will use this approach in the next two sections for BBU Hoteling and C-RAN systems.

\subsection{Gaussian Approximation}
\label{gaus:error}
Throughout this paper, we adopt the Gaussian approximation proposed in \cite{wei}, based on the work in \cite{poly}, to evaluate the probability $\mathrm{P_{e}}(r,k,\textbf{H})$ of decoding error for a transmission at rate $r$ in a slot of $k$ channel uses when the channel matrix is $\textbf{H}$. This amounts to
\begin{equation}
\mathrm{P_{e}}(r,k,\textbf{H})=Q\left(\frac{C(\textbf{H})-r}{\sqrt{\frac{V(\textbf{H})}{k}}}\right),\label{error-TI-MIMO}
\end{equation}
where we have defined
\begin{equation}
\begin{split} & C(\textbf{H})=\sum_{j=1}^{m_{rt}}\log_{2}\left(1+\frac{s\lambda_{j}}{m_{t}}\right)\\
&\mathrm{\textrm{and}}~V(\textbf{H})=\left(m_{rt}-\sum_{j=1}^{m_{rt}}\frac{1}{\left(1+\frac{s\lambda_{j}}{m_{t}}\right)^{2}}\right)\log_2^2e,
\end{split}
\label{MI-pe}
\end{equation}
with $m_{rt}=\textrm{min}(m_{r},m_{t})$; $\{\lambda_{j}\}_{j=1,...,m_{rt}}$
being the eigenvalues of the matrix $\textbf{H}^{H}\textbf{H}$; and
$Q(\cdot)$ being the Gaussian complementary cumulative distribution
function. Expressions obtained by
means of the Gaussian approximation (\ref{error-TI-MIMO})
will be marked for simplicity of notation as equalities in the following.

For future reference, we note that we have the limit
\begin{equation}\label{ref:shan}
\underset{k\longrightarrow\infty}{\lim} \mathrm{P_{e}}(r,k,\textbf{H})=\left\{\begin{array}{ll}
1&\mathrm{if~}C(\textbf{H})<r\\
0&\mathrm{if~}C(\textbf{H})>r
\end{array}\right.
\end{equation}
in the asymptotic regime of large blocklengths.
\subsection{Average Latency}
The comparison between the latency of the conventional BBU Hoteling and C-RAN implementations of HARQ and the approach proposed in this paper, which is based on the separation of data and control planes, depends on the specific fronthaul transport latency in the system of interest. To elaborate, we define as $L_f$ the two-way latency for fronthaul transport and processing at the BBU as measured in terms of number of transmission slots. Furthermore, we neglect for simplicity the time required to transmit ACK/NACK messages in the downlink, although this could be easily included as a common term in all latency expressions. The overall average latency $D_c$ of the conventional HARQ implementation in BBU Hoteling and C-RAN is then given as
\begin{equation}\label{convdelay}
D_c=\mathrm{E}[N](1+L_f),
\end{equation}
which is measured in terms of number of transmission slots, where we recall that $N$ is the number of retransmissions of the HARQ protocol. The average latency (\ref{convdelay}) follows since each retransmission requires one time slot for uplink transmission and $L_f$ time slots for two-way fronthaul transmission and BBU processing. In contrast, the average latency of the proposed approach can be approximated as
\begin{equation}\label{prodelay}
D_s=\mathrm{E}[N],
\end{equation}
since each retransmission can be immediately acknowledged by the RRHs without having to wait for feedback from the BBU. In this regard, we also recall that the processing needed at the RRHs with local feedback is minimal, since it does not entail any decoding, and hence the corresponding latency is much smaller than the processing time needed at the BBU to decode the data packet. From (\ref{convdelay}) and (\ref{prodelay}), the ratio of the average latencies for the two implementations is $D_c/S_s=1+L_f$.

While our work is not tied to a specific standard or system, current standardization efforts and industry white papers have reported the two-way fronthaul latency $L_f$ to consist of a two-way fronthaul transport latency of around $0.5$ ms for single-hop fronthaul links \cite{ngm} and of a BBU processing time of around $2.3$-$2.6$ ms \cite[p.38]{ero}. As a result, the two-latency $L_f$, for time slots of duration $1$ ms, is no smaller than $3$ and potentially much larger, e.g., in the presence of a multihop fronthaul architecture (see also \cite{gulati}). As a result, the ratio $D_c/D_s$ can be of the order of $4$ or larger, showing the significant latency reduction achievable via the proposed approach.

\section{Analysis of RRH-Based Low-Latency Local Feedback for BBU Hoteling}
\label{dran:an}
In this section, we analyze the performance in terms of throughput and probability of success of the low-latency local feedback scheme for BBU Hoteling as introduced in Sec. \ref{dran}. We focus separately on the three standard modes of HARQ-TI, CC and IR, in order of complexity \cite{harq-comp}. We recall that, in the considered low-latency scheme, a decision to stop retransmissions is made by the RRH by sending an ACK message, while a retransmission is decided by the transmission of a NAK message. We define as $\mathrm{ACK}_n$ the event that an ACK message is sent at the $n$th transmission attempt and as $\mathrm{NAK}_n$ the event that a NAK message is sent for all the first $n$ transmissions. Therefore, in applying the analytical expression introduced in the previous section, we can focus on the evaluation of the probabilities $\mathrm{P}(\mathrm{RTX}_n)=\mathrm{P}(\mathrm{NAK}_n)$ and $\mathrm{P}(\mathrm{D}_{n}|\mathrm{STOP}_{n})=\mathrm{P}(\mathrm{D}_{n}|\mathrm{ACK}_{n})$ in order to calculate throughput and probability of success. Throughout, we use the Gaussian approximation for the probability of error discussed in Sec. \ref{gaus:error}.
\subsection{HARQ-TI}
\label{sec:TI}
With HARQ-TI, the same packet is retransmitted by the
UE upon reception of a NAK message until the maximum number $n_{max}$ of
retransmissions is reached or until an ACK message is received. Moreover, decoding at the
BBU is based on the last received packet only. HARQ-TI is hence a standard ARQ strategy \cite{arq}.

\subsubsection{Ideal Feedback}
For reference, we first study the ideal case in which zero-delay feedback is available directly from BBU. Using the approximation (\ref{error-TI-MIMO}) and averaging over the channel distribution, the approximate probability
of an erroneous decoding at the BBU at the $n$th retransmission is given by $\mathrm{E}\left[\mathrm{P}_{e}(r,k,\textbf{H}_n)\right]$. Accordingly, since with HARQ-TI the BBU performs decoding independently for each slot, we obtain
\begin{equation}
\mathrm{P}(\mathrm{NAK}_{n})=\left(\mathrm{E}\left[\mathrm{P}_{e}(r,k,\textbf{H})\right]\right)^{n}.
\end{equation}
As discussed, throughput and the probability of success now can be calculated as (2)-(4) and (5), where the throughput can be simplified as
\begin{equation}\label{t:val}
T=r\left(1-\mathrm{E}\left[\mathrm{P}_{e}(r,k,\textbf{H})\right]\right).
\end{equation}
The average in (\ref{t:val}) can be computed numerically based on the known distribution of the eigenvalues the Wishart-distributed matrix $\textbf{H}^H\textbf{H}$, see \cite[Theorem 2.17]{matrix}. As an important special case, for a SISO link ($m_t=m_r=1$), we have $|H|^2\sim\mathcal{X}^2_2$ and hence
\begin{equation}
\mathrm{E}\left[\mathrm{P}_{e}(r,k,H)\right]=\int_0^\infty\mathrm{P}_{e}(r,k,\sqrt{x})f_{\mathcal{X}^2_2}(x)\mathrm{d}x.
\end{equation}

\subsubsection{Local Feedback}
With local feedback, as discussed, at each transmission attempt $n$, the RRH estimates the current channel
realization $\textbf{H}_{n}$ and decides whether it expects the BBU
to decode correctly or not by comparing the probability of error by using the following rule (\ref{rule_gen}), which reduces to
\begin{equation}
\begin{split}\mathrm{P_{e}}(r,k,\textbf{H}_{n})\underset{\mathrm{NAK}}{\overset{\mathrm{ACK}}{\lessgtr}}\mathrm{P_{th}},
\end{split}
\label{pe}
\end{equation}
since decoding is done only based on the last received packet. We
observe that, in the case of a single antenna at the transmitter and/or
the receiver, the rule (\ref{pe}) only requires the RRH to estimate
the SNR $s||\textbf{H}_{n}||^{2}/m_{t}$.

The quantities that are needed to calculate the performance metrics under study can be then directly obtained from their definitions as
\begin{equation}\label{p_decod}
\mathrm{P}(\mathrm{D}_{n}|\mathrm{ACK}_{n})=1-\mathrm{E}\left[\mathrm{P_{e}}(r,k,\textbf{H})|\mathrm{P_{e}}(r,k,\textbf{H})\leq\mathrm{P_{th}}\right]
\end{equation}
\begin{equation}
\begin{split}\mathrm{and~}\mathrm{P}(\mathrm{NAK}_{n}) & =\left(\mathrm{P}\left(\mathrm{P_{e}}(r,k,\textbf{H})>\mathrm{P_{th}}\right)\right)^{n}.\end{split}
\label{nak_ack_2}
\end{equation}
As discussed, (\ref{p_decod}) and (\ref{nak_ack_2}) can be obtained by averaging over the distribution of the eigenvalues of $\textbf{H}^H\textbf{H}$. As an example, for a SISO link, we obtain
\begin{alignat}{1}
&\mathrm{P}(\mathrm{D}_{n}|\mathrm{ACK}_{n})=\notag\\
&1-\frac{1}{1-F_{\mathcal{X}^2_2}(\gamma(\mathrm{P_{th}}))}\int^\infty_{\gamma(\mathrm{P_{th}})}\mathrm{P_{e}}(r,k,\sqrt{x})f_{\mathcal{X}^2_2}(x)\mathrm{d}x\\
&\mathrm{and}~\mathrm{P}(\mathrm{NAK}_{n})  =\left(F_{\mathcal{X}^2_2}(\gamma(\mathrm{P_{th}}))\right)^{n},
\end{alignat}
where $\gamma(\mathrm{P_{th}})$ is calculated by solving the non-linear equation
\begin{equation}\label{gamma}
\mathrm{P_{e}}\left(r,k,\sqrt{\gamma(\mathrm{P_{th}}})\right)=\mathrm{P_{th}},
\end{equation}
e.g., by means of bisection.

%and hence it can be bounded as
%\begin{equation}
%\begin{split}T & =\\
% & \frac{r\sum_{n=1}^{N}\mathrm{P}(\mathrm{D}_{n}|\mathrm{ACK}_{n})\mathrm{P}(\mathrm{ACK}_{n})}{\sum_{n=1}^{n_{m}-1}n\left(\mathrm{P}(\mathrm{ACK}_{n})\right)+n_{m}\mathrm{P}(\mathrm{NAK}_{n_{m}-1})},
%\end{split}
%\label{thro_low22}
%\end{equation}
%where $\mathrm{P}(\mathrm{NAK}_{n})$ is given in (\ref{nak_ack_2}), $\mathrm{P}(\mathrm{D}_{n}|\mathrm{ACK}_{n})$ is given by (\ref{p_decod}), $\mathrm{P}(\mathrm{ACK}_{n})$ can be calculated by (\ref{p_ack}) and we have used the fact that $\sum_{n=1}^{n_{m}}\mathrm{P}(\mathrm{ACK}_{n})=1-\mathrm{P}(\mathrm{NAK}_{n_{max}})$.
%
%\textbf{Remark:} As we mentioned before, the throughput of TI for the ideal feedback scheme does not deponed on the value of $n_{max}$. Also, since the maximum throughput of local feedback is smaller-equal to the ideal feedback, the optimum threshold for local feedback can be obtained by setting $\mathrm{P_{th}}=1$. But setting $\mathrm{P_{th}}=1$ does not enforce  the probability of success to be maximized. Hence, instead of throughput, we consider the probability of success to evaluate the performance of TI scheme in Sec. \ref{sec:sim}.

\subsection{HARQ-CC}
\label{sec:cc}
With HARQ-CC, every retransmission of the UE consists
of the same encoded packet as for TI. However, at the $n$th transmission attempt,
 the BBU uses maximum ratio combining (MRC) of all the $n$ received
packets in order to improve the decoding performance. For HARQ-CC, we only consider here a SISO link. This is because MRC requires to compute
the weighted sum of the received signals across multiple transmission attempts, where the weight
is given by the corresponding scalar channel for a SISO link. Note that SIMO and MISO links
could also be tackled in a similar way by considering weights obtained from the effective scalar
channels, although we do not explicitly consider these cases in the paper. Due to MRC, at the $n$th retransmission, the received signal can be written as
\begin{equation}
\bar{y}_{n}=\frac{\sum_{i=1}^{n}H^{*}_{i}y_i}{\bar{S}_{n}},
\end{equation}
or equivalently as
\begin{equation}
\bar{y}_{n}=\bar{S}_{n}x+\bar{w}_{n},
\end{equation}
where $y_{n}$ is the $n$th received packet, the noise $\bar{w}_{n}$ is distributed as  $\mathcal{CN}(0,1)$ and the effective channel gain of the combined signal is given by $\bar{S}_{n}=\sqrt{\sum_{i=1}^{n}|{H}_{i}|^{2}}$.

\subsubsection{Ideal Feedback}
The probability that
the BBU does not decode correctly when the effective SNR is $\bar{S}_{n}^2$
is given as $\mathrm{P_e}(r,k,\bar{S}_n)$. Let $\bar{\mathrm{D}}_n$ denote the event that the $n$th transmission is not decoded correctly at the BBU. The probability of the event $\mathrm{NAK}_{n}$ is then given as $\mathrm{P}(\mathrm{NAK}_{n})=\mathrm{P}(\bigcap_{j=1}^n\bar{\mathrm{D}}_j)$, which can be upper bounded, using the chain rule of probability, as
\begin{equation}\label{argu1}
\begin{split}
\mathrm{P}(\mathrm{NAK}_{n})&=\mathrm{P}\left(\bar{\mathrm{D}}_n\right)\mathrm{P}\left(\bar{\mathrm{D}}_{n-1}|\bar{\mathrm{D}}_n\right)\cdots\mathrm{P}\left(\bar{\mathrm{D}}_1|\bigcap^n_{j=2}\bar{\mathrm{D}}_j\right)\\
&\leq \mathrm{P}\left(\bar{\mathrm{D}}_n\right)=\mathrm{E}\left[\mathrm{P_e}(r,k,\bar{S}_n)\right].
\end{split}
\end{equation}
The usefulness of the bound (\ref{argu1}) for small values of k will be validated in Sec. \ref{sec:sim} by means of a comparison with Monte Carlo simulations. We also refer to \cite{sassi} where the same bound is proposed as an accurate approximation of the probability of error for HARQ-CC. We note that the inequality (\ref{argu1}) is asymptotically tight in the limit of a large blocklength, since the limit $\mathrm{P}(\bar{\mathrm{D}}_m|\bigcap_{j=m+1}^n\bar{\mathrm{D}}_j)\rightarrow1$ as $k\rightarrow\infty$ holds for a fixed $r$ due to (\ref{ref:shan}) and to the inequality $\bar{S}_n\geq\bar{S}_m$ for $n\geq m$. The usefulness of the bound (\ref{argu1}) for small values of $k$ will be validated in Sec. \ref{sec:sim} by means of a comparison with Monte Carlo simulations.
%\begin{equation}\label{argu}
%\mathrm{P}(\bar{\mathrm{D}}_n)=\mathrm{P_{e}}(r,k,\bar{S}_{n})\leq\mathrm{P_{e}}(r,k,\bar{S}_{m})=\mathrm{P}(\bar{\mathrm{D}}_m).
%\end{equation}
%According to the coding theorem, if the transmission rate is smaller than capacity, the probability of error can be made smaller by increasing the blocklength and if the transmission rate is larger than capacity, the probability of error goes to one as blocklength increases. Therefore, since for any $n>m$ the effective SNR of the $n$th transmission $\bar{S}^2_n$ is equal grater than the effective SNR of $m$th transmission $\bar{S}^2_m$ for any $n>m$, all the conditional probabilities go to $1$ as the blocklength goes to infinity. Hence, the upper in (\ref{argu1}) is accurate for large blocklength codes and it provides an approximation for small blocklengths. For a Rayleigh fading channel,
Since the effective SNR is distributed as $\bar{S}^2_{n}=\sum_{i=1}^{n}|{H}_{i}|^{2}\sim\mathcal{X}^2_{2n}$, the bound (\ref{argu1}) can be calculated as
\begin{equation}\label{nak:cc2}
\mathrm{P}(\mathrm{NAK}_{n})\leq \int_0^\infty\mathrm{P_{e}}(r,k,\sqrt{x})f_{\mathcal{X}^2_{2n}}(x)\mathrm{d}x.
\end{equation}

\subsubsection{Local Feedback}
With local feedback, the RRH decision is made according to the rule $\mathrm{P_{e}}(r,k,\bar{S}_{n})\lessgtr_\mathrm{NAK}^\mathrm{ACK}\mathrm{P_{th}}$, for a threshold $\mathrm{P_{th}}$ to be optimized. Similar to (\ref{p_decod}) and (\ref{nak_ack_2}), we can compute the probabilities
\begin{equation}\label{first:D}
\begin{split}
\mathrm{P}(\mathrm{D}_{n}|\mathrm{ACK}_{n})&=1-\mathrm{E}\Big[\mathrm{P_{e}}(r,k,\bar{S}_{n})|\{\mathrm{P_{e}}(r,k,\bar{S}_{n-1})>\mathrm{P_{th}}\}\\
&\bigcap\{\mathrm{P_{e}}(r,k,\bar{S}_{n})\leq\mathrm{P_{th}}\}\Big]
\end{split}
\end{equation}
\begin{equation}\label{:d}
\mathrm{and~} \mathrm{P}(\mathrm{NAK}_{n})=\mathrm{P}[\mathrm{P_{e}}(r,k,\bar{S}_{n})>\mathrm{P_{th}}].
\end{equation}
Note that in (\ref{first:D})-(\ref{:d}) we used the fact that, if the condition $\mathrm{P_{e}}(r,k,\bar{S}_{n})>\mathrm{P_{th}}$ holds, then we also have the inequality $\mathrm{P_{e}}(r,k,\bar{S}_{i})>\mathrm{P_{th}}$ for all the indices $i<n$ due to the monotonicity of the probability $\mathrm{P_{e}}(r,k,\bar{S})$ as a function of $\bar{S}$. Furthermore, noting that we can write $\bar{S}^2_{n}=\bar{S}^2_{n-1}+|H_n|^2$, where $\bar{S}^2_{n-1}\sim\mathcal{X}^2_{2n-2}$ and $|H_n|^2\sim\mathcal{X}^2_2$ are independent, from (\ref{first:D}) and (\ref{:d}), we have
\begin{equation}
\begin{split}
&\mathrm{P}(\mathrm{D}_{n}|\mathrm{ACK}_{n})=1-\mathrm{E}\Big[\mathrm{P_{e}}(r,k,\bar{S}_{n})|\left\{\bar{S}^2_{n-1}<\gamma(\mathrm{P_{th}})\right\}\\
&~~~~~~~~~~~~~~~~~~~~\bigcap\left\{\bar{S}^2_{n-1}+|H_n|^2\geq\gamma(\mathrm{P_{th}})\right\}\Big]\\
&=1-\frac{1}{\Delta(\gamma(\mathrm{P_{th}}))}\int_{0}^{\gamma(\mathrm{P_{th}})}\int_{\gamma(\mathrm{P_{th}})-y}^{\infty}\mathrm{P_{e}}(r,k,\sqrt{x+y})\\
&~~~~~~~~~~~~~~~~~~~~~~~~~~~~~~~~~~~~~f_{\mathcal{X}^2_2}(x)f_{\mathcal{X}^2_{2n-2}}(y)\mathrm{d}x\mathrm{d}y\\
&\mathrm{and}~\mathrm{P}(\mathrm{NAK}_{n})=F_{\mathcal{X}^2_{2n}}(\gamma(\mathrm{P_{th}})),
%&=\int_0^{\gamma^*(\mathrm{P_{th}})}\mathrm{P_{e}}(r,k,\sqrt{\lambda_n})f(\lambda_n)d\lambda_n,
\end{split}
\end{equation}
where $\Delta(\gamma(\mathrm{P_{th}}))$ is defined as
\begin{equation}
\begin{split}
&\Delta(\gamma(\mathrm{P_{th}}))=\int_{0}^{\gamma(\mathrm{P_{th}})}\int_{\gamma(\mathrm{P_{th}})-y}^\infty f_{\mathcal{X}^2_2}(x)f_{\mathcal{X}^2_{2n-2}}(y)\mathrm{d}x\mathrm{d}y.
\end{split}
\end{equation}

\subsection{HARQ-IR}
\label{sec:ir} With HARQ-IR, the UE
transmits new parity bits at each transmission attempt and the BBU performs
decoding based on all the received packets.

\subsubsection{Ideal Feedback}
With HARQ-IR,  a set of $n$ transmission attempt for a given information messages
can be treated as the transmission over $n$ parallel channels (see, e.g., \cite{caire}),
and hence the error probability at the $n$th transmission can be
computed as $\mathrm{P_{e}}(r,k,\mathcal{H}_{n})$ %=Q\left(\frac{C(\mathcal{H}_{n})-r}{\sqrt{\frac{V(\mathcal{H}_{n})}{k}}}\right),\label{pe_mimo_ir}
where $\mathcal{H}_{n}=\mathrm{diag}([\textbf{H}_{1},...,\textbf{H}_{n}])$ \cite{wei}. Moreover, following the same argument as (\ref{argu1}), the decoding error at the $n$th transmission can be upper bounded as
\begin{equation}\label{nak-ideal-ir}
\mathrm{P}(\mathrm{NAK}_{n})\leq \mathrm{P}(\bar{\mathrm{D}}_n)=\mathrm{E}[\mathrm{P_{e}}(r,k,\mathcal{H}_{n})],
\end{equation}
which is tight for large values of $k$ due to (\ref{ref:shan}). This can be computed using the known distribution of the eigenvalues of the matrices $\textbf{H}_{i}^H\textbf{H}_{i}$ and the independence of the matrices $\textbf{H}_i$ for $i=1,...,n$. For instance in the SISO case, we get
\begin{equation}\label{nak-me}
\begin{split}
\mathrm{P}&(\mathrm{NAK}_{n})\leq\int_0^\infty\cdots \int_0^\infty\mathrm{P_{e}}(r,k,\mathrm{diag}([\sqrt{x_1},...,\sqrt{x_n}]))\\
&~~~~~~~~~~~~~~~~~~\prod_{i=1}^n f_{\mathcal{X}^2_2}(x_i)\mathrm{d}x_1\cdots \mathrm{d}x_n.
\end{split}
\end{equation}

%\begin{figure}
%\centerline{\epsfig{figure=estimation_lossfig.eps,scale=.6}}\caption{Throughput loss of the ideal feedback caused by the estimate in (\ref{argu1}) versus blocklength for HARQ-CC and HARQ-IR ($m_t=1$, $m_r=1$, $n_{max}=5$ for $r=1$ bits/symbol and $r=3$ bits/symbol).}
%\label{thro_loss}
%\end{figure}
\subsubsection{Local Feedback}
\label{IR-dran-local}
With local feedback, at the $n$th retransmission, the RRH sends feedback
to the UE according to the rule $\mathrm{P_{e}}(r,k,\mathcal{H}_{n})\lessgtr_\mathrm{NAK}^\mathrm{ACK}\mathrm{P_{th}}$. Due to the monotonicity
of the probability $\mathrm{P_{e}}(r,k,\mathcal{H}_{n})$ as a function of each eigenvalue, we have that the probability $\mathrm{P_{e}}(r,k,\mathcal{H}_{n})$ is no larger than $\mathrm{P_{e}}(r,k,\mathcal{H}_{n-1})$. Therefore, similar to CC, we can calculate
\begin{alignat}{1} &\mathrm{P}(\mathrm{D}_{n}|\mathrm{ACK}_{n})=1-\mathrm{E}[\mathrm{P_{e}}(r,k,\mathcal{H}_{n})|\mathcal{A}(\mathrm{P_{th}})]\label{pdick}\\
&\mathrm{and~}\mathrm{P}(\mathrm{NAK}_{n})=\mathrm{P}(\mathrm{P_{e}}(r,k,\mathcal{H}_{n})>\mathrm{P_{th}})\label{pnick},
\end{alignat}
where we have defined the event $\mathcal{A}(\mathrm{P_{th}})=\{\{\mathrm{P_{e}}(r,k,\mathcal{H}_{n-1})>\mathrm{P_{th}}\}\bigcap\{\mathrm{P_{e}}(r,k,\mathcal{H}_{n})\leq\mathrm{P_{th}}\}\}$.
For the SISO case, we can calculate these quantities as
\begin{equation}\label{int_IR}
\begin{split}
&\mathrm{P}(\mathrm{D}_{n}|\mathrm{ACK}_{n})=1\\
&~~~~~~~~-\frac{1}{\Delta(\mathrm{P_{th}})}\int_0^{\infty}\cdots \int_0^{\infty}\mathrm{P_{e}}(r,k,\mathrm{diag}([\sqrt{x_1},...,\sqrt{x_n}]))\\
&~~~~~~~~\textbf{1}\left(\mathcal{A}(\mathrm{P_{th}})\right)\prod_{i=1}^n f_{\mathcal{X}^2_2}(x_i)\mathrm{d}x_1\cdots \mathrm{d}x_n\\
&\mathrm{and}~\mathrm{P}(\mathrm{NAK}_{n})=\\
&\int_0^{\infty}\cdots \int_0^{\infty}\textbf{1}\left(\mathrm{P_{e}}(r,k,\mathrm{diag}([\sqrt{x_1},...,\sqrt{x_n}]))>\mathrm{P_{th}}\right)\\
&~~~~~~~~~~~~~~\prod_{i=1}^n f_{\mathcal{X}^2_2}(x_i)\mathrm{d}x_1\cdots \mathrm{d}x_n,
\end{split}
\end{equation}
where
\begin{equation}
\begin{split}
&\Delta(\mathrm{P_{th}})=\int_0^{\infty}\cdots \int_0^{\infty}\textbf{1}\left(\mathcal{A}(\mathrm{P_{th}})\right)\prod_{i=1}^n f_{\mathcal{X}^2_2}(x_i)\mathrm{d}x_1\cdots \mathrm{d}x_n.
\end{split}
\end{equation}

\section{Analysis of User-Centric Low-Latency Local Feedback for C-RAN}
\label{cran:an}
In this section, we turn to the analysis of the user centric low-latency local feedback schemes introduced in Sec. \ref{cran} for C-RAN. Throughout, we focus on HARQ-IR for its practical relevance, see, e.g., \cite{lte}. Furthermore, we consider the case where each RRH has only one receiving antenna, i.e., $m_{r,l}=1$ for $l=1,..,L$. Extensions to other HARQ protocols and to scenarios with large number of antennas at the RRHs are possible by following similar arguments as in the previous sections and will not be further discussed here. We recall that in a C-RAN with local feedback, the retransmission decisions are made at the UE based on feedback from the RRHs. We treat separately the case of ideal zero-delay feedback from the BBU, and the hard and soft feedback schemes in the following.

%In , it is proposed that if the decoding error probability of the received signal at a RRH is smaller than a threshold $\mathrm{P_{th}}$, an ACK feedback should be sent to the UE, otherwise, a NAK should be transmitted back.

\subsection{Ideal Feedback}
\label{sec:ideal}
We first consider for reference the case of zero-delay ideal feedback from the BBU. Since the BBU jointly processes all the received signals for decoding, at the $n$th retransmission, the signal available at the BBU can be written, using (\ref{fist-sig}), as $\textbf{y}^n=[\textbf{y}^T_1,...,\textbf{y}^T_n]^T$, where
\begin{equation}
\textbf{y}_n=\sqrt{\frac{s}{m_{t}}}\textbf{H}_{n}\textbf{x}_n+\textbf{w}_{n},
\end{equation}
with  $\textbf{H}_{n}=[\textbf{h}^T_{1,n} ~\textbf{h}^T_{2,n}\cdots \textbf{h}^T_{L,n}]^T$ and $\textbf{w}_{n}=[w_{1,n}~ w_{2,n}\cdots w_{L,n}]^T$. We emphasize  that we denoted here as $\textbf{h}_{l,n}$ instead of $\textbf{H}_{l,n}$ the vector containing the channel coefficients between the UE and $l$th RRH in the $n$th retransmission, so as to stress the focus on single-antenna RRHs.  The effective received signal is hence given by
\begin{equation}
\textbf{y}^n=\sqrt{\frac{s}{m_{t}}}\mathcal{H}_n[\textbf{x}_1^T \cdots  \textbf{x}_n^T]^T+[\textbf{w}^T_{1}\cdots \textbf{w}^T_{n}]^T,
\end{equation}
with $\mathcal{H}_{n}=\mathrm{diag}([\textbf{H}_{1},...,\textbf{H}_{n}])$. Therefore, the decoding error probability at the $n$th transmission is given by $\mathrm{P}_{\mathrm{e}}(r,k,\mathcal{H}_{n})$.

The C-RAN performance in terms of throughput and the probability of success under ideal feedback can be obtained following the discussion in Sec. \ref{per-cert} by computing the probability $\mathrm{P}(\mathrm{RTX}_n)$ that a retransmission is required at the $n$th transmission attempt. This can be bounded similar to (\ref{nak-ideal-ir}) as $\mathrm{P}(\mathrm{RTX}_n)\leq \mathrm{P}(\bar{\mathrm{D}}_n)=\mathrm{E}[\mathrm{P_{e}}(r,k,\mathcal{H}_{n})]$.

\subsection{Hard Feedback Scheme}
\label{sec:hard}
With the hard feedback low-latency scheme described in Sec. \ref{hard:detail}, each RRH calculates its own decoding error probability  $\mathrm{P}_{\mathrm{e}}(r,k,\mathcal{H}_{l,n})$ with $\mathcal{H}_{l,n}=\mathrm{diag}(\textbf{h}_{l,1} ~\textbf{h}_{l,2}\cdots \textbf{h}_{l,n})$ and uses the rule (\ref{rule_hard}), which reduces to
\begin{equation}
\mathrm{P_{e}}(r,k,\mathcal{H}_{l,n})\underset{\mathrm{NAK}}{\overset{\mathrm{ACK}}{\lessgtr}}\mathrm{P_{th}}.
\end{equation}
%and quantizes this value by mapping it to $0$ if $\mathrm{P}_{\mathrm{e}}(r,k,\mathcal{H}_{l,n})<\mathrm{P}_\mathrm{th}$ or to $1$ otherwise.
Each RRH sends a single bit indicating the ACK/NAK feedback to the UE. The UE decides that a retransmission is necessary as long as all the RRHs return a NAK message, and it stops retransmission otherwise.

Throughput and probability of success can be computed as detailed in Sec. \ref{per-cert} by using the following probabilities
\begin{alignat}{1} &\mathrm{P}(\mathrm{D}_{n}|\mathrm{STOP}_{n})=1\notag\\
&-\mathrm{E}\left[\mathrm{P_{e}}(r,k,\mathcal{H}_{n})\middle|\prod_{l=1}^L\textbf{1}\left(\mathrm{P}_{\mathrm{e}}(r,k,\mathcal{H}_{l,n})>\mathrm{P}_\mathrm{th}\right)=0\right]\\
&\mathrm{and~}\mathrm{P}\left(\mathrm{RTX}_{n}\right)=\mathrm{P}\left(\prod_{l=1}^L\textbf{1}\left(\mathrm{P}_{\mathrm{e}}(r,k,\mathcal{H}_{l,n})>\mathrm{P}_\mathrm{th}\right)=1\right).
\end{alignat}
The above probabilities can be calculated similar to the equations derived in Sec. \ref{dran:an} by averaging over the distribution of the eigenvalues of the involved channel matrices.

\begin{figure}
\centerline{\epsfig{figure=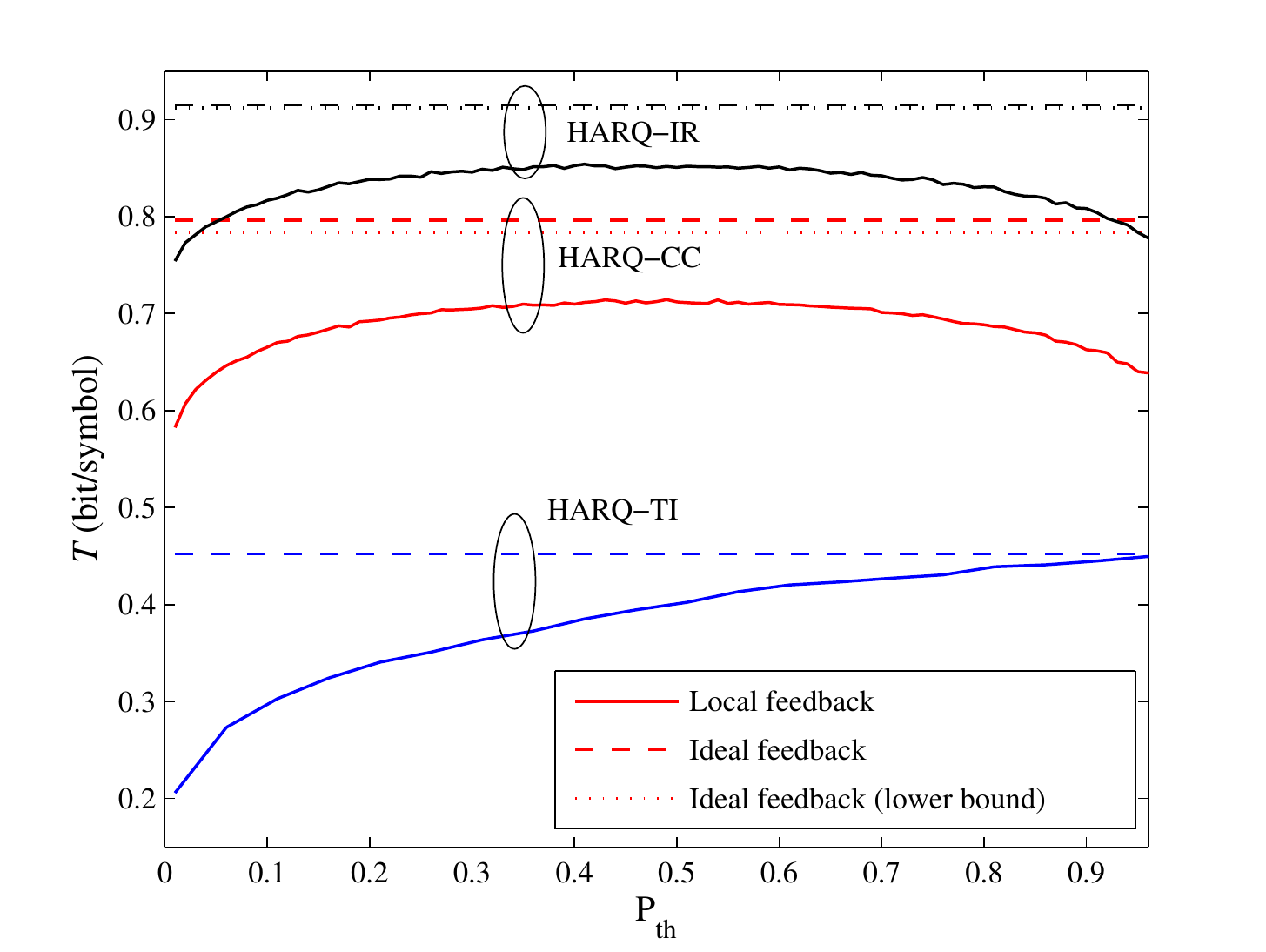,scale=.66}} \protect\caption{Throughput versus threshold $\mathrm{P_{th}}$ for ideal feedback and local feedback in a BBU Hoteling system ($s=3$ dB, $n_{max}=5$, $r=2$ bit/symbol, $k=50$, $m_t=1$ and $m_r=1$).}
\label{r-pth-5}
\end{figure}
%We use a uniform quantizer that quantizes $\mathrm{P}_{\mathrm{e}}(r,k,\mathcal{H}_{n})$ to $R_{up}$ if  $R_{low}<\mathrm{P}_{\mathrm{e}}(r,k,\mathcal{H}_{n})\leq R_{up}$. The probability $\hat{\mathrm{P}}_{\mathrm{e}}(r,k,\mathcal{H}_{n})$ in (\ref{pe:UE1}) can be interpreted as the decoding error probability at the BBU if the BBU decodes the signals received from each RRH separately.

\subsection{Soft Feedback Scheme}
\label{sec:soft}
%(\ref{soft:detail})
With the soft feedback introduced in Sec. \ref{cran}, each RRH quantizes the local CSI $\textbf{h}_{l,n}$ with $b$ bits. From the $b$ feedback bits received from each RRH, the UE obtains the quantized channel vectors $\hat{\textbf{h}}_{l,n}$ for $l\in\{1,...,L\}$. Based of these, the decision (\ref{rule-soft}) is adopted, which reduces to
\begin{equation}
\mathrm{P_{e}}(r,k,\hat{\mathcal{H}}_{n})\underset{\mathrm{RTX}}{\overset{\mathrm{STOP}}{\lessgtr}}\mathrm{P_{th}},
\end{equation}
where $\hat{\mathcal{H}}_n=\mathrm{diag}(\hat{\textbf{H}}_{1},...,\hat{\textbf{H}}_{n})$ and $\hat{\textbf{H}}_{n}=[\hat{\textbf{h}}^T_{1,n}\cdots \hat{\textbf{h}}^T_{2,n}\cdots \hat{\textbf{h}}^T_{L,n}]^T$ collect the quantized CSI. Accordingly, we can compute the desired probabilities as
\begin{alignat}{1} &\mathrm{P}(\mathrm{D}_{n}|\mathrm{STOP}_{n})=1-\mathrm{E}[\mathrm{P_{e}}(r,k,\mathcal{H}_{n})|\mathrm{P}_{\mathrm{e}}(r,k,\hat{\mathcal{H}}_{n})\leq\mathrm{P_{th}}]\\
&\mathrm{and~}\mathrm{P}(\mathrm{RTX}_{n})=\mathrm{P}(\mathrm{P}_{\mathrm{e}}(r,k,\hat{\mathcal{H}}_{n})>\mathrm{P_{th}}).
\end{alignat}
The above probabilities can be computed analytically or via Monte Carlo simulations by averaging over the distribution of the eigenvalues similar to Sec. \ref{dran:an}.

%\section{Joint BBU-edge HARQ Control}
%
%In the previous sections, we have assumed that HARQ control plane is fully implemented at the edge of the network, that is, either at the RRHs for BBU Hoteling or at the RRHs and UEs for C-RAN systems. This approach reflects a scenario in which feedback from the BBUs is not used to drive the HARQ process due to the two-way fronthaul latency. Nevertheless, when feedback from the BBUs can be received with some delay $n_{fh}$ that is smaller than the maximum number $n_{max}$ of transmission attempts, this information can be be used to possibly modify the HARQ decisions.
%
%In this section, we briefly consider this possibility by returning to the BBU Hoteling model studied in Sec. \ref{sec:ir} and by considering a generalized system model in which feedback from the BBU has a delay $n_{fh}$ with $n_{fh} \in{0,1,\cdots,n_{max}}$. Note that $n_{fh}=0$ corresponds to ideal feedback, $n_{fh}=1$ to a single block of delay, and so on, until $n_{fh}=n_{max}$, which corresponds to local feedback. We propose that, if the feedback received from the BBU is positive, the RRH send an ACK message irrespective of the outcome of the decision rule (\ref{????}). Instead, if a NAK message is received, the protocol continues as discussed in Sec. ?.
%
%Throughput and probability of success of this scheme can be obtained from ... by ...
\section{Numerical Results and Discussion}
\label{sec:sim}

In this section, we validate the analysis presented in the previous sections and provide insights on the performance comparison of ideal and local feedback schemes for BBU Hoteling and C-RAN systems via numerical examples.
\begin{figure}
\centerline{\epsfig{figure=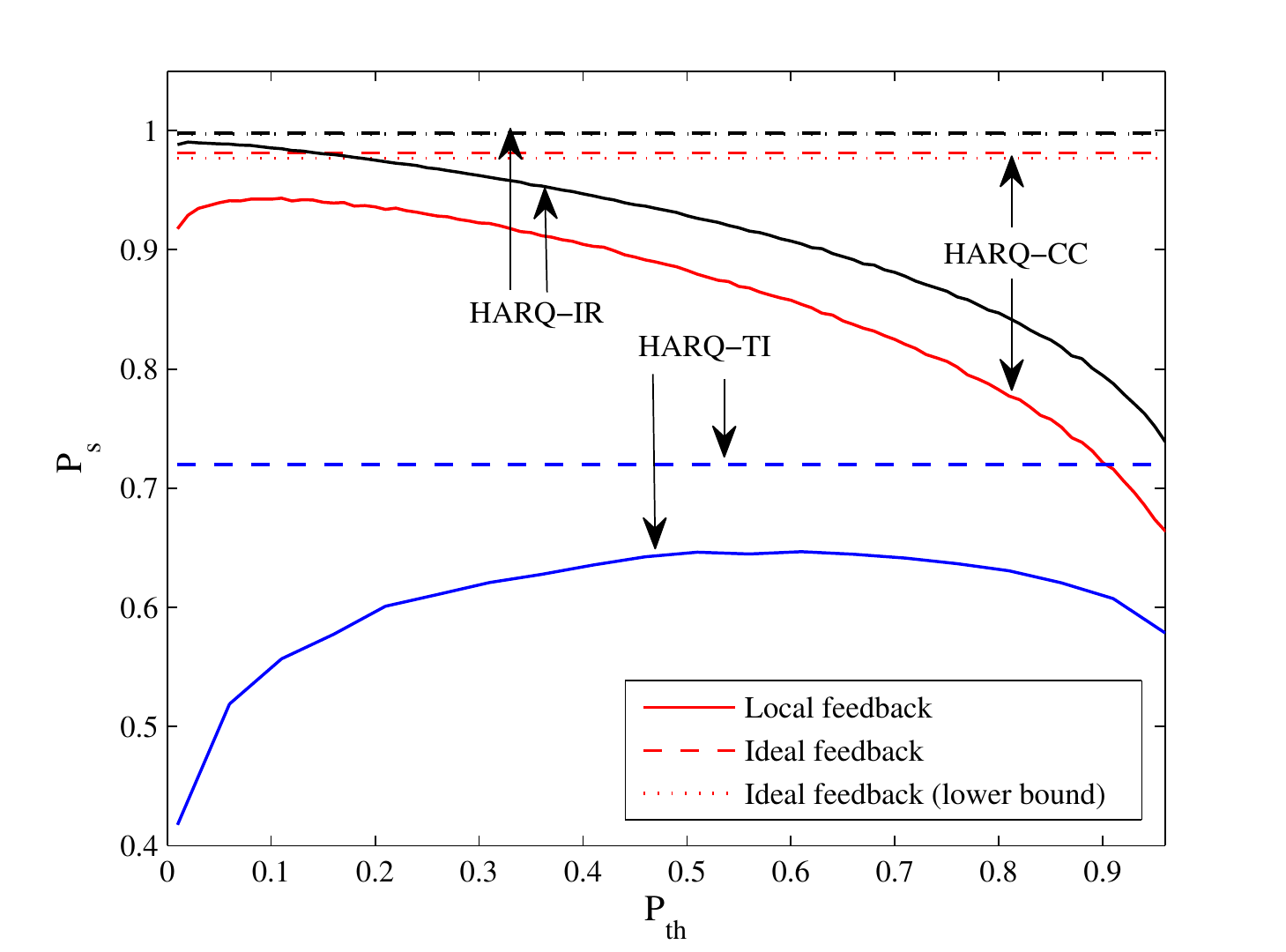,scale=.66}}\caption{Probability of success versus threshold $\mathrm{P_{th}}$ in a BBU Hoteling system ($s=3$ dB, $n_{max}=5$, $r=2$ bit/symbol, $k=50$, $m_t=1$ and $m_r=1$).}
\label{ps_pth}
\end{figure}

\subsection{BBU Hoteling}
We first study the optimization of the threshold $\mathrm{P_{th}}$ used in the local feedback schemes. As an exemplifying case study, we consider the case of BBU Hoteling described in Sec. \ref{dran:an}. In Fig. \ref{r-pth-5} and Fig. \ref{ps_pth}, respectively,
the throughput $T$ and the probability of success $\mathrm{P_s}$ are shown versus
$\mathrm{P_{th}}$ for $s=3$ dB, $n_{max}=5$ retransmissions, $r=2$ bit/symbol and blocklength $k=50$
for a SISO link, i.e., for $m_t=1$ and $m_r=1$. The curves have been computed using both the equations derived in Sec. \ref{dran:an} and Monte Carlo simulations. The latter refer to the simulation of the HARQ process in which the probability of error at the BBU is modeled by means of the Gaussian approximation. The analytical results are confirmed to match with the Monte Carlo simulations, except for the ideal feedback performance of HARQ-CC and HARQ-IR, for which, as discussed in Sec. \ref{dran:an}, the expressions (\ref{nak:cc2}) and (\ref{nak-me}) yield lower bounds on throughput and probability of success. As seen in the figures, the bounds are very accurate for $k$ as small as $50$.

From Fig. \ref{r-pth-5} and Fig. \ref{ps_pth}, it is also concluded that throughput and probability of success are maximized for different values of threshold $\mathrm{P}_\mathrm{th}$, with the throughput metric requiring a larger threshold. In fact, a larger value of $\mathrm{P_{th}}$, while possibly causing the acknowledgement of packets that will be incorrectly decoded at the BBU, may enhance the throughput by allowing for the transmission of fresh information in a new HARQ session. This is particularly evident for HARQ-TI, for which setting $\mathrm{P_{th}}=1$ guarantees a throughput equal to the case of ideal feedback, but at the cost of a loss in the probability of success. It is also observed that more powerful HARQ schemes such as CC and IR are more robust to a suboptimal choice of $\mathrm{P_{th}}$ in terms of throughput, although lower values of $\mathrm{P_{th}}$  are necessary in order to enhance the probability of success by avoiding a premature transmission of an ACK message.

\begin{figure}
\centerline{\epsfig{figure=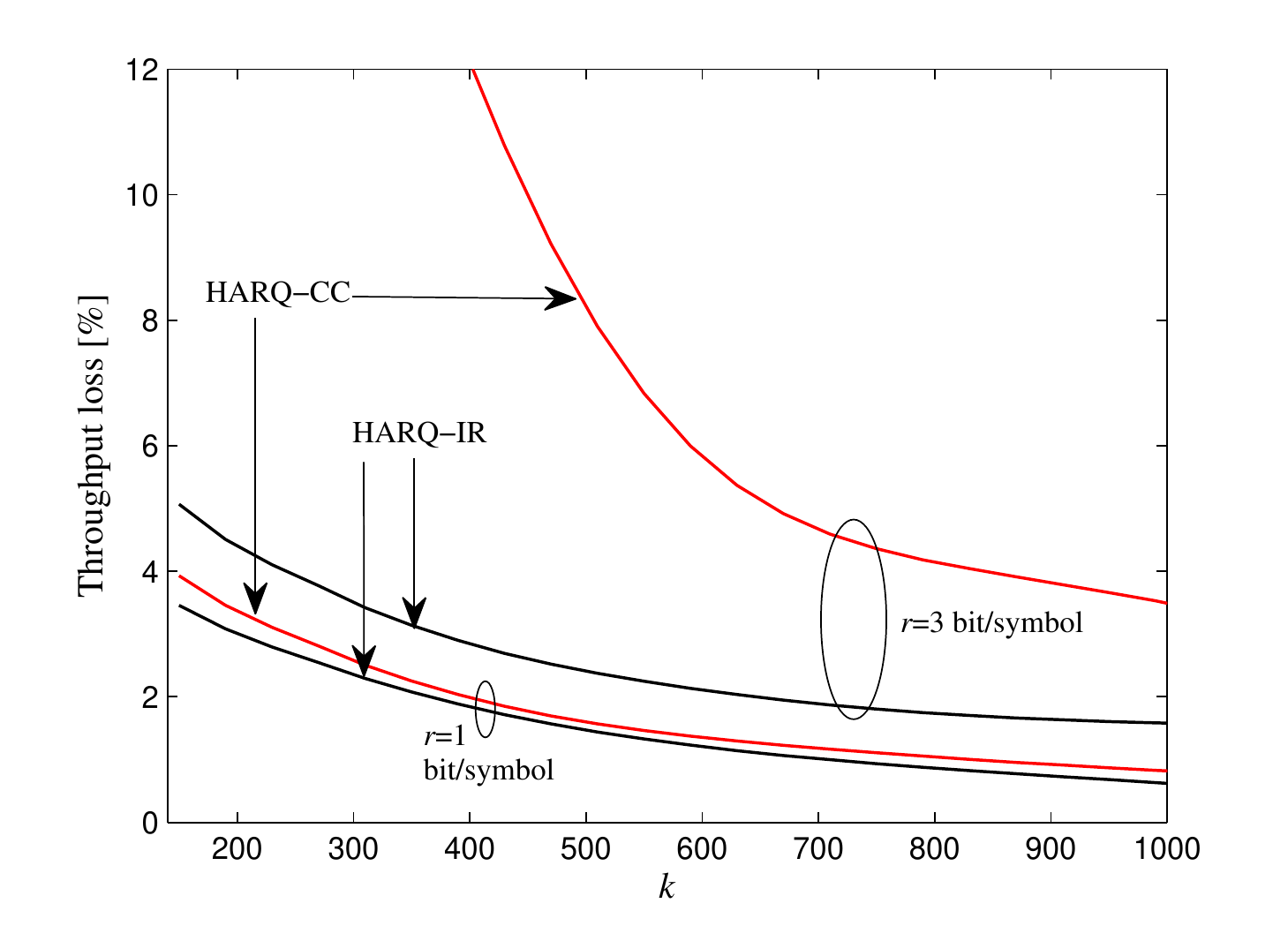,scale=.66}}\caption{Throughput loss versus blocklength $k$ for HARQ-CC and HARQ-IR in a BBU Hoteling system ($s=4$ dB, $n_{max}=10$ $m_t=1$, $m_r=1$, $\mathrm{P_s}>0.99$ for $r=1$ bit/symbol and $r=3$ bit/symbol).}
\label{g-k-N}
\end{figure}

We now illustrate in Fig. \ref{g-k-N} the throughput loss of local feedback as compared to the ideal feedback case, as a function of the blocklength $k$, for two rates $r=1$ bit/symbol and $r=3$ bit/symbol for HARQ-CC and HARQ-IR in a BBU Hoteling system. Henceforth, to avoid clutter in the figures, we only show Monte Carlo results, given the match with analysis discussed above. The simulation are performed by setting $s=4$ dB, $n_{max}=10$ and we focus on a SISO link.
For every value of $k$, the threshold $\mathrm{P_{th}}$ is optimized
to maximize the throughput $T$ under the constraint that the probability of success satisfies the requirement $\mathrm{P_s}>0.99$ (see, e.g., \cite{lte} and \cite{yester}). It can be seen that, as the blocklength
increases, the performance loss of local feedback decreases significantly. This reflects a fundamental insight: The
performance loss of local feedback is due to the fact that the local
decisions are taken by the RRH based only on channel state information, without
reference to the specific channel noise realization that affects the
received packet. Therefore, as the blocklength $k$ increases, and hence
as the errors due to atypical channel noise realizations become less
likely, the local decisions tend to be consistent with the actual
decoding outcomes at the BBU. In other words, as the blocklength $k$ grows larger, it becomes easier for the RRH to predict the decoding outcome at the BBU: In the Shannon regime of infinite $k$, successful or unsuccessful  decoding depends deterministically  on wether the rate $r$ is above or below capacity.

\begin{figure}
\centerline{\epsfig{figure=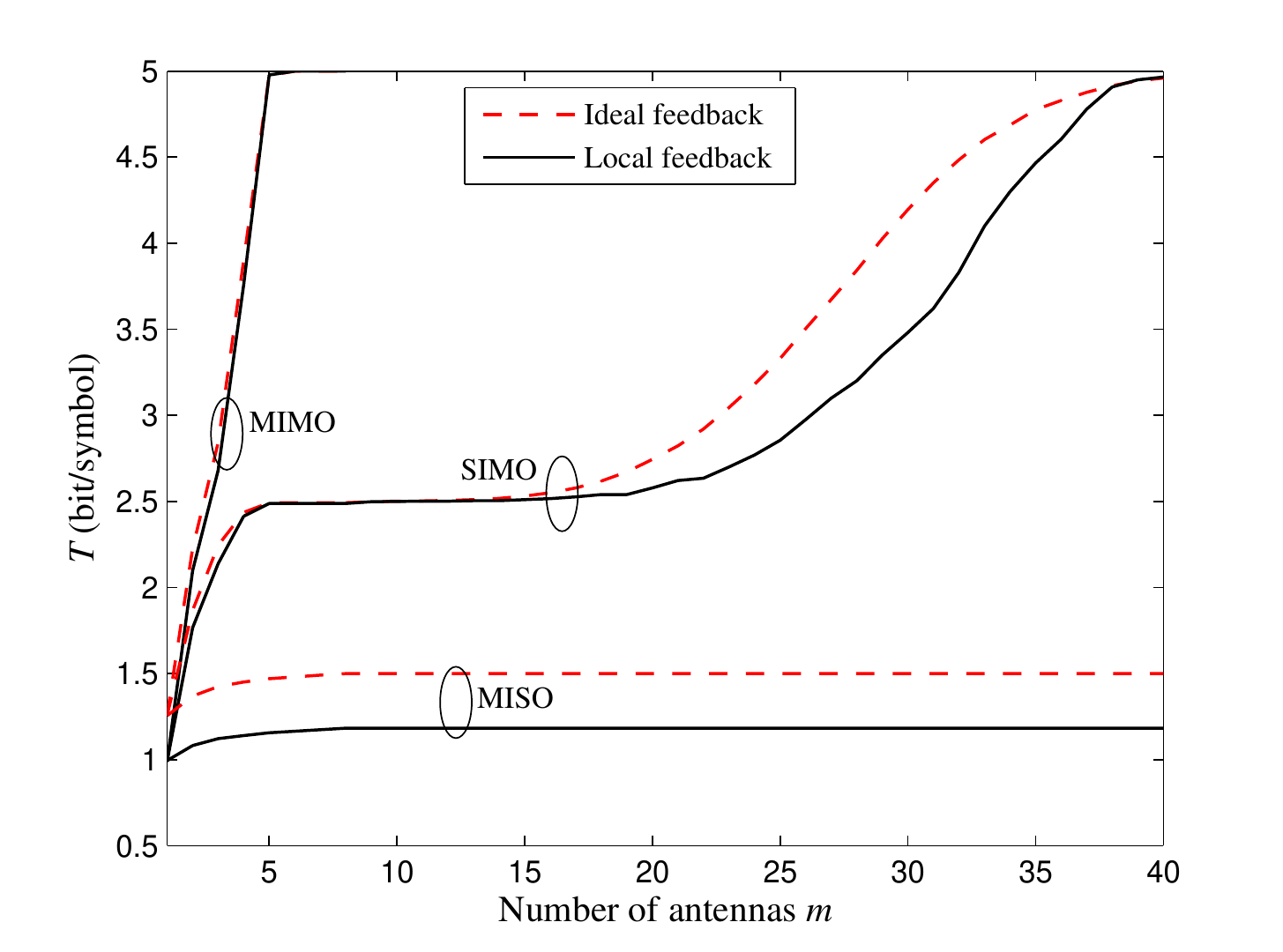,scale=.66}}\caption{Throughput versus the number of antennas for MISO, SIMO and MIMO with HARQ-IR in a BBU Hoteling system ($s=1$ dB, $n_{max}=10$, $r=5$ bit/symbol, $k=100$ and $\mathrm{P_s}>0.99$).}
\label{TI-comp}
\end{figure}
A related conclusion can be reached from Fig.  \ref{TI-comp}, where we investigate the throughput for MIMO ($m_t=m_r=m$), MISO ($m_t=m$ and $m_r=1$) and SIMO ($m_t=1$ and $m_r=m$) links versus the number of antennas $m$ for HARQ-IR, with $s=1$ dB, $n_{max}=10$, $r=5$ bit/symbol, $k=100$. As in Fig. \ref{g-k-N}, the threshold $\mathrm{P}_{\mathrm{th}}$ is optimized here, and henceforth, to maximize the throughput under the constraint $\mathrm{P_s}>0.99$. As $m$ grows large, it is seen that the throughput of SIMO and MIMO increases up to the maximum throughput $T=r=5$ bit/symbol. This is unlike the case with MISO, since an increase in the number of transmit antennas only enhances the diversity order but does not improve the average received SNR, yielding a ceiling on the achievable throughput that is smaller than the maximum throughput $T=5$ bit/symbol. We remark that the interest in large values of $m$ stems from massive MIMO systems. We also note that the flattening of the throughput for SIMO around $T = 2.5$ bit/symbol for $m$ between $6$ and around $20$ antennas is due to the fact that, for the given range of $m$, at least two retransmissions are necessary, which implies a throughput equal to $T=r/2=2.5$ bit/symbol (see also Fig. \ref{versSNR} for related discussion). As for the throughput loss caused by local feedback, we observe that it is generally minor, ranging from at most $0.27$ bit/symbol for MIMO to at most $0.73$ bit/symbol for MISO.

\subsection{C-RAN}
We now turn our attention to the performance of low-latency local feedback for HARQ-IR over C-RAN systems with $L>1$ single-antenna RRHs and $m_t=4$ antennas at the UE. Throughout, we consider the throughput of local feedback based on hard or soft feedback, under the constraint $\mathrm{P_s}>0.99$ on the probability of success. As a reference, we also consider the performance of a BBU Hoteling system, i.e., with $L=1$, under both ideal and local feedback (we mark the latter as ``hard feedback'' following the discussion in Sec. \ref{sec:hard}).

For soft feedback, we set different values for the number of feedback bits $b$, including $b=\infty$, with the latter being equivalent to a BBU Hoteling system with three co-located antennas at the RRH (i.e., $m_{r,1}=3$ and $L=1$). We use a vector quantizer for each RRH $l$, in which $b'\leqslant b$ bits are used to quantize the channel direction $\textbf{h}_{l,n}/||\textbf{h}_{l,n}||$ and $b-b'$ bits for the amplitude $||\textbf{h}_{l,n}||$. For vector quantization, we generate randomly quantization codebooks with normalized columns (see, e.g., \cite{quant}) until finding one for which the constraint on the probability of success is met. The amplitude $||\textbf{h}_{l,n}||$ of each channel vector is quantized with the remaining $b'-b$ using a quantizer with numerically optimized thresholds. For $b=3$, $b=6$, $b=9$ and $b=16$, the number of bits used for the quantization of the direction of each channel vector are $b'=1$, $b'=4$, $b'=5$ and $b'=12$.

\begin{figure}
\centerline{\epsfig{figure=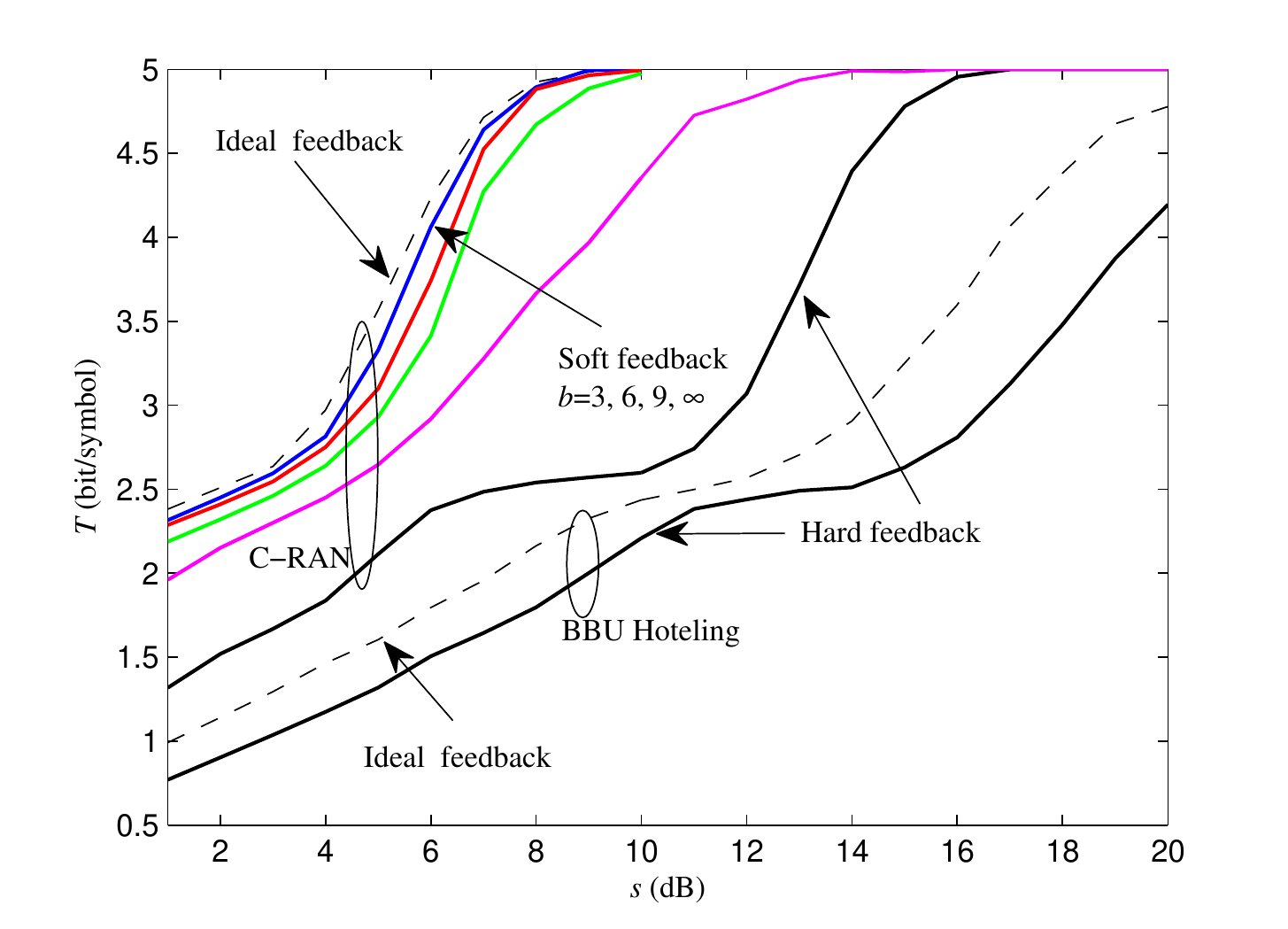,scale=.66}}\caption{Throughput versus SNR $s$ for BBU Hoteling ($L=1$) and C-RAN ($L=3$) systems ($n_{max}=10$, $r=5$ bit/symbol, $k=100$, $\mathrm{P_s}>0.99$, $m_t=4$, $m_{r,l}=1$).}
\label{versSNR}
\end{figure}
In Fig. \ref{versSNR}, the throughput of the schemes outlined above is shown versus the SNR parameter $s$. We first observe that hard feedback, which only require $1$ bit of feedback per RRH, is able to improve over the performance of BBU Hoteling, but the throughput is limited by the errors due to the user-centric local decisions based on partial feedback from the RRHs. This limitation is partly overcome by implementing the soft feedback scheme, whose throughput increases for a growing feedback rate. Note that, even with an infinite feedback rate, the performance of local feedback still exhibits a gap as compared to ideal feedback for the same reasons discussed above for BBU Hoteling systems. Also, the flattening of the throughput of less performing schemes around $T=2.5$ for intermediate SNR levels is due to the need to carry out at least two retransmissions unless the SNR is sufficiently large (see, e.g., \cite{ir-perform}).

%The throughput of BBU Hoteling for SNRs $s\in[10 ,12]$ is $2.4$ bit/symbol which implies that approximately $2$ retransmission is required to successfully decode a packet. The throughput of BBU Hoteling does not change significantly in this range due to the fact that the increase in SNR does not decrease the required number of transmission. Also, the hard feedback shows the same behavior. For SNRs $s>13$, the throughput gets larger than $2.5$ bit/symbol implying that the received packet is decodable even with one transmission. This trend in HARQ-IR is explained in \cite{ir-perform}.

\begin{figure}
\centerline{\epsfig{figure=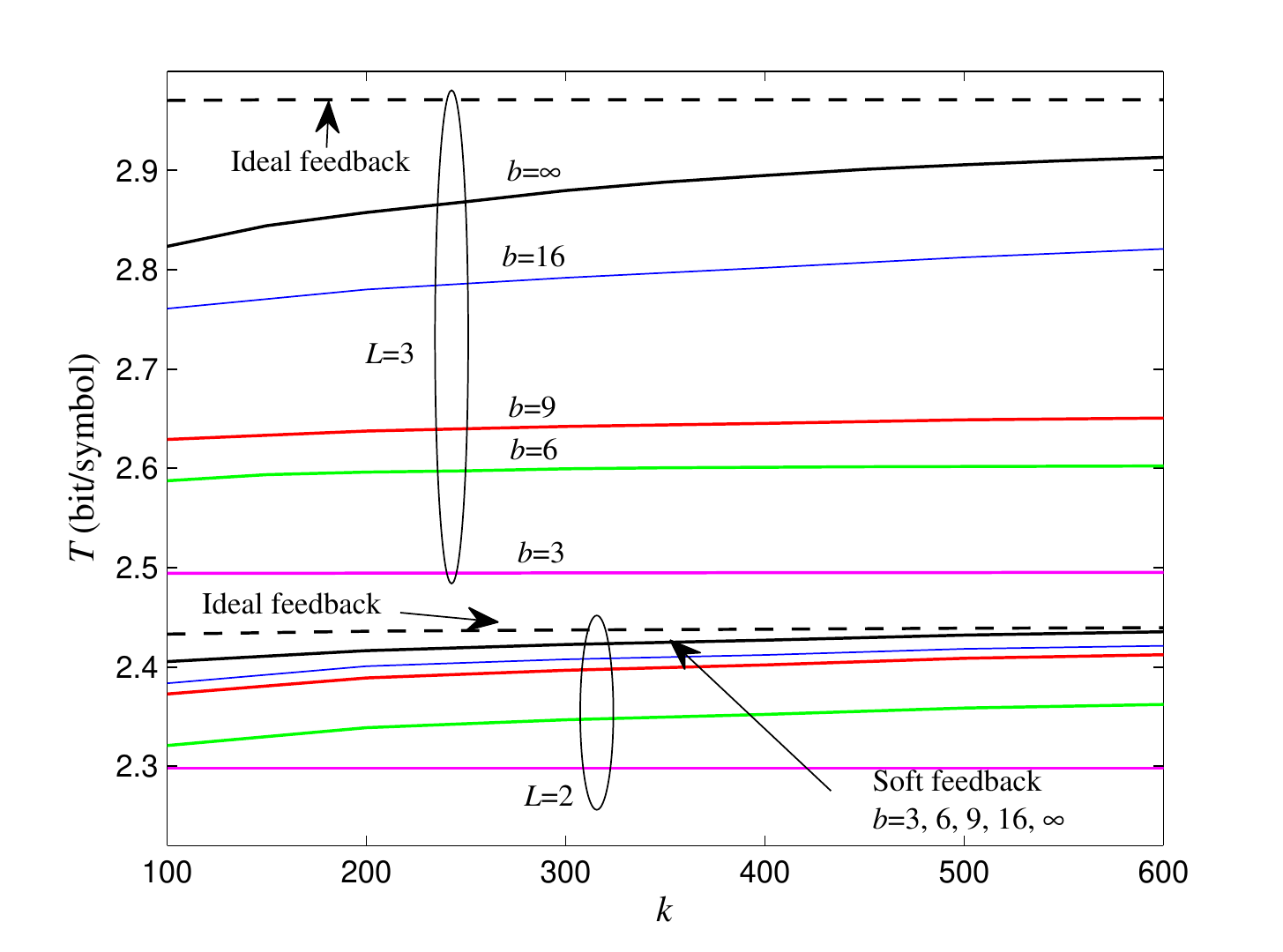,scale=.66}}\caption{Throughput versus blocklength $k$ for ideal and soft feedback schemes in C-RAN with $L=3$ and $L=2$. The throughput of the hard feedback scheme for $L=2$ and $L=3$ (not shown) are $T=1.7$ and $T=1.77$, respectively ($s=4$ dB, $n_{max}=10$, $r=5$ bit/symbol, $\mathrm{P_s}>0.99$, $m_t=4$, $m_{r,l}=1$).}
\label{versk}
\end{figure}

We finally show in Fig. \ref{versk} the throughput of ideal and soft  feedback schemes versus the blocklength $k$ for a C-RAN system with $L=2$ and $L=3$. We observe that, in a C-RAN system with a sufficiently small feedback rate such as $b=3$ and $b=6$, an increase in the blocklength $k$ does not significantly increase the throughput, which is limited by the CSI quantization error. However, with a larger $b$, such as $b=16$, the throughput can be more significantly improved towards the performance of ideal feedback, especially for a smaller number of RRHs.
%Moreover, the loss between the ideal feedback and the soft schemes in the presence of $2$ RRHs is much smaller compared to the scenario with $3$ RRHs. This can be explained by noting that as the number of RRHs grows, since the effect of noise is less pronounced, the throughput of the ideal feedback, and similarly, the soft feedback schemes are increased. However, this increase in the number of RRHs results in larger quantization error due to the fact that the UE receives more quantized channel vectors to make a centric low latency local decision which causes a larger loss between the soft and ideal feedback strategies for larger values of $L$.

\section{Concluding Remarks}
\label{sec:con}
The performance of BBU Hoteling and C-RAN systems is currently under close scrutiny as limitations due to constraints imposed by fronthaul capacity and latency are increasingly brought to light (see, e.g., \cite{c-ran}). An important enabling technology to bridge the gap between the desired lower cost and higher spectral efficiency of BBU Hoteling and C-RAN and its potentially poor performance in terms of throughput at higher layers is the recently proposed control and data separation architecture \cite{moham}. In this context,  this work has considered BBU Hoteling and C-RAN systems in which retransmission decisions are made at the edge of the network, that is, by the RRHs or UEs, while data decoding is carried out in a centralized fashion at the BBUs.

As shown, for BBU Hoteling, this class of solutions has the potential to yield throughput values close to those achievable with ideal zero-delay feedback from the BBUs, particularly when the packet is sufficiently long or the number of received antennas is large enough. For C-RAN, it was argued that multi-bit feedback messages from the RRHs are called for in order to reduce the throughput loss and a specific scheme based on vector quantization was proposed to this end.

Interesting future work include the analysis of control and data separation architectures for C-RAN systems for the purpose of user detection activity in random access in scenarios with a massive number of devices.

\bibliographystyle{IEEEtran}
\bibliography{refrences}

\begin{thebibliography}{10}
\providecommand{\url}[1]{#1}
\csname url@samestyle\endcsname
\providecommand{\newblock}{\relax}
\providecommand{\bibinfo}[2]{#2}
\providecommand{\BIBentrySTDinterwordspacing}{\spaceskip=0pt\relax}
\providecommand{\BIBentryALTinterwordstretchfactor}{4}
\providecommand{\BIBentryALTinterwordspacing}{\spaceskip=\fontdimen2\font plus
\BIBentryALTinterwordstretchfactor\fontdimen3\font minus
  \fontdimen4\font\relax}
\providecommand{\BIBforeignlanguage}[2]{{%
\expandafter\ifx\csname l@#1\endcsname\relax
\typeout{** WARNING: IEEEtran.bst: No hyphenation pattern has been}%
\typeout{** loaded for the language `#1'. Using the pattern for}%
\typeout{** the default language instead.}%
\else
\language=\csname l@#1\endcsname
\fi
#2}}
\providecommand{\BIBdecl}{\relax}
\BIBdecl

\bibitem{itashah}
S.~Khalili and O.~Simeone, ``Control-data separation in cloud {RAN}: The case
  of uplink {HARQ},'' in \emph{Proc. Information Theory and Applications
  (ITA)}, La Jolla, Jan. 2016.

\bibitem{hotel}
\BIBentryALTinterwordspacing
{5G Public Private Partnership}, ``{5G PPP 5G} architecture.'' [Online].
  Available: \url{https://5g-ppp.eu/white-papers/}
\BIBentrySTDinterwordspacing

\bibitem{c-ran}
\BIBentryALTinterwordspacing
{China Mobile Research Institute}, ``{C-RAN}: {The} road towards green {RAN},
  white paper,'' 2010. [Online]. Available:
  \url{http://labs.chinamobile.com/cran/wp-content/uploads/CRAN\_white\_paper\_v2\_5\_EN.pdf.}
\BIBentrySTDinterwordspacing

\bibitem{check}
A.~Checko, H.~Christiansen, Y.~Yan, L.~Scolari, G.~Kardaras, M.~Berger, and
  L.~Dittmann, ``Cloud {RAN} for mobile networks-a technology overview,''
  \emph{IEEE Commun. Surveys Tuts.}, vol.~17, no.~1, pp. 405--426, First
  quarter 2015.

\bibitem{nahas}
M.~Nahas, A.~Saadani, J.~Charles, and Z.~El-Bazzal, ``Base stations evolution:
  Toward {4G} technology,'' in \emph{Proc. Int. Conf. on Telecommunications
  (ICT)}, pp. 1-6, Aalborg, Denmark, Apr. 2012.

\bibitem{ngm}
\BIBentryALTinterwordspacing
{NGMN Alliance}, ``Further study on critical {C-RAN} technologies, {White}
  paper,'' 2015. [Online]. Available:
  \url{https://www.ngmn.org/uploads/media/NGMN\_RANEV\_D2\_Further\_Study\_on\_Critical\_C-RAN\_Technologes\_v1.0.pdf.}
\BIBentrySTDinterwordspacing

\bibitem{arq}
A.~Cipriano, P.~Gagneur, G.~Vivier, and S.~Sezginer, ``Overview of {ARQ} and
  {HARQ} in beyond {3G} systems,'' in \emph{Proc. {IEEE} Int. Symp. on
  Personal, Indoor and Mobile Radio Communications (PIMRC)}, pp. 424-429,
  Istanbul, Turkey, Sep. 2010.

\bibitem{rrh-bell}
U.~D{\"o}tsch, M.~Doll, H.-P. Mayer, F.~Schaich, J.~Segel, and P.~Sehier,
  ``Quantitative analysis of split base station processing and determination of
  advantageous architectures for {LTE},'' \emph{Bell Labs Technical Journal},
  vol.~18, no.~1, pp. 105--128, Jun. 2013.

\bibitem{gulati}
S.~Gulati, B.~Natarajan, S.~Kalyanasundaram, and R.~Agrawal, ``Performance
  analysis of centralized {RAN} deployment with non-ideal fronthaul in
  {LTE}-advanced networks,'' in \emph{Proc. IEEE Vehicular Technology
  Conference (VTC Spring)}, pp. 1--5, Nanjing, China May 2016.

\bibitem{osiarc}
\BIBentryALTinterwordspacing
Q.~Han, C.~Wang, M.~Levorato, and O.~Simeone, ``On the effect of fronthaul
  latency on {ARQ} in {C-RAN} systems.'' [Online]. Available:
  \url{http://arxiv.org/abs/1510.07176}
\BIBentrySTDinterwordspacing

\bibitem{rost1}
D.~Wubben, P.~Rost, J.~Bartelt, M.~Lalam, V.~Savin, M.~Gorgoglione, A.~Dekorsy,
  and G.~Fettweis, ``Benefits and impact of cloud computing on {5G} signal
  processing: Flexible centralization through {Cloud}-{RAN},'' \emph{IEEE
  Signal Process. Mag.}, vol.~31, no.~6, pp. 35--44, Nov. 2014.

\bibitem{olive}
A.~De~La~Oliva, X.~Costa~Perez, A.~Azcorra, A.~Di~Giglio, F.~Cavaliere,
  D.~Tiegelbekkers, J.~Lessmann, T.~Haustein, A.~Mourad, and P.~Iovanna,
  ``{Xhaul}: toward an integrated fronthaul/backhaul architecture in {5G}
  networks,'' \emph{IEEE Wireless Commun.}, vol.~22, no.~5, pp. 32--40, Oct.
  2015.

\bibitem{moham}
A.~Mohamed, O.~Onireti, M.~Imran, A.~Imran, and R.~Tafazolli, ``Control-data
  separation architecture for cellular radio access networks: A survey and
  outlook,'' \emph{to appear in IEEE Commun. Surveys Tuts.}, 2015.

\bibitem{rost}
P.~Rost and A.~Prasad, ``Opportunistic hybrid {ARQ}-enabler of
  centralized-{RAN} over nonideal backhaul,'' \emph{IEEE Wireless Commun.
  Letters}, vol.~3, no.~5, pp. 481--484, Oct. 2014.

\bibitem{poly}
Y.~Polyanskiy, H.~Poor, and S.~Verd\'{u}, ``Channel coding rate in the finite
  blocklength regime,'' \emph{IEEE Trans. Inf. Theory}, vol.~56, no.~5, pp.
  2307--2359, May 2010.

\bibitem{yuri}
Y.~Polyanskiy, ``Channel coding: non-asymptotic fundamental limits.''\hskip 1em
  plus 0.5em minus 0.4em\relax Ph.D. thesis, Princeton university, 2010.

\bibitem{lte}
{E. Dahlman, S. Parkvall, J. Skold, P. Bemin}, \emph{{3G} Evolution: {HSPA} and
  {LTE} for Mobile Broadband}.\hskip 1em plus 0.5em minus 0.4em\relax Academic
  Press, second edition, 2008.

\bibitem{quant}
D.~Love, R.~Heath, and T.~Strohmer, ``Grassmannian beamforming for
  multiple-input multiple-output wireless systems,'' \emph{IEEE Trans. Inf.
  Theory}, vol.~49, no.~10, pp. 2735--2747, Oct. 2003.

\bibitem{caire}
G.~Caire and D.~Tuninetti, ``The throughput of hybrid-{ARQ} protocols for the
  {Gaussian} collision channel,'' \emph{IEEE Trans. Inf. Theory}, vol.~47,
  no.~5, pp. 1971--1988, Jul. 2001.

\bibitem{wei}
\BIBentryALTinterwordspacing
W.~Yang, G.~Durisi, T.~Koch, and Y.~Polyanskiy, ``Quasi-static {MIMO} fading
  channels at finite blocklength.'' [Online]. Available:
  \url{http://arxiv.org/abs/1311.2012.}
\BIBentrySTDinterwordspacing

\bibitem{ero}
N.~Nikaein, ``Processing radio access network functions in the cloud: Critical
  issues and modeling,'' in \emph{in Proc. Int. Workshop Mobile Cloud
  Computing, Services (MCS)}, pp. 36-42, Paris, France Sep. 2015.

\bibitem{harq-comp}
P.~Frenger, S.~Parkvall, and E.~Dahlman, ``Performance comparison of {HARQ}
  with chase combining and incremental redundancy for {HSDPA},'' in \emph{Proc.
  IEEE Vehicular Technology Conference (VTC)}, vol.~3, pp. 1829--1833, Atlantic
  City, New Jersey, USA, Oct. 2001.

\bibitem{matrix}
A.~M. Tulino and S.~Verdu, \emph{Random Matrix Theory and Wireless
  Communications}.\hskip 1em plus 0.5em minus 0.4em\relax Now Publishers Inc,
  2004.

\bibitem{sassi}
\BIBentryALTinterwordspacing
{R. Sassioui, E. Pierre-Doray, L. Szczecinski, B. Pelletier}, ``Modelling
  decoding errors in {HARQ}.'' [Online]. Available:
  \url{http://arxiv.org/abs/1512.02511}
\BIBentrySTDinterwordspacing

\bibitem{yester}
A.~Lozano and N.~Jindal, ``Are yesterday's information-theoretic fading models
  and performance metrics adequate for the analysis of today's wireless
  systems?'' \emph{IEEE Commun. Mag.}, vol.~50, no.~11, pp. 210--217, Nov.
  2012.

\bibitem{ir-perform}
I.~Stanojev, O.~Simeone, and Y.~Bar-Ness, ``Performance analysis of
  collaborative hybrid-{ARQ} incremental redundancy protocols over fading
  channels,'' in \emph{Proc. IEEE Signal Processing Advances in Wireless
  Communications (SPAWC)}, pp. 1--5, Cannes, France, Jul. 2006.

\end{thebibliography}
\begin{IEEEbiography}[{\includegraphics[width=1in,height=1.25in,clip,keepaspectratio]{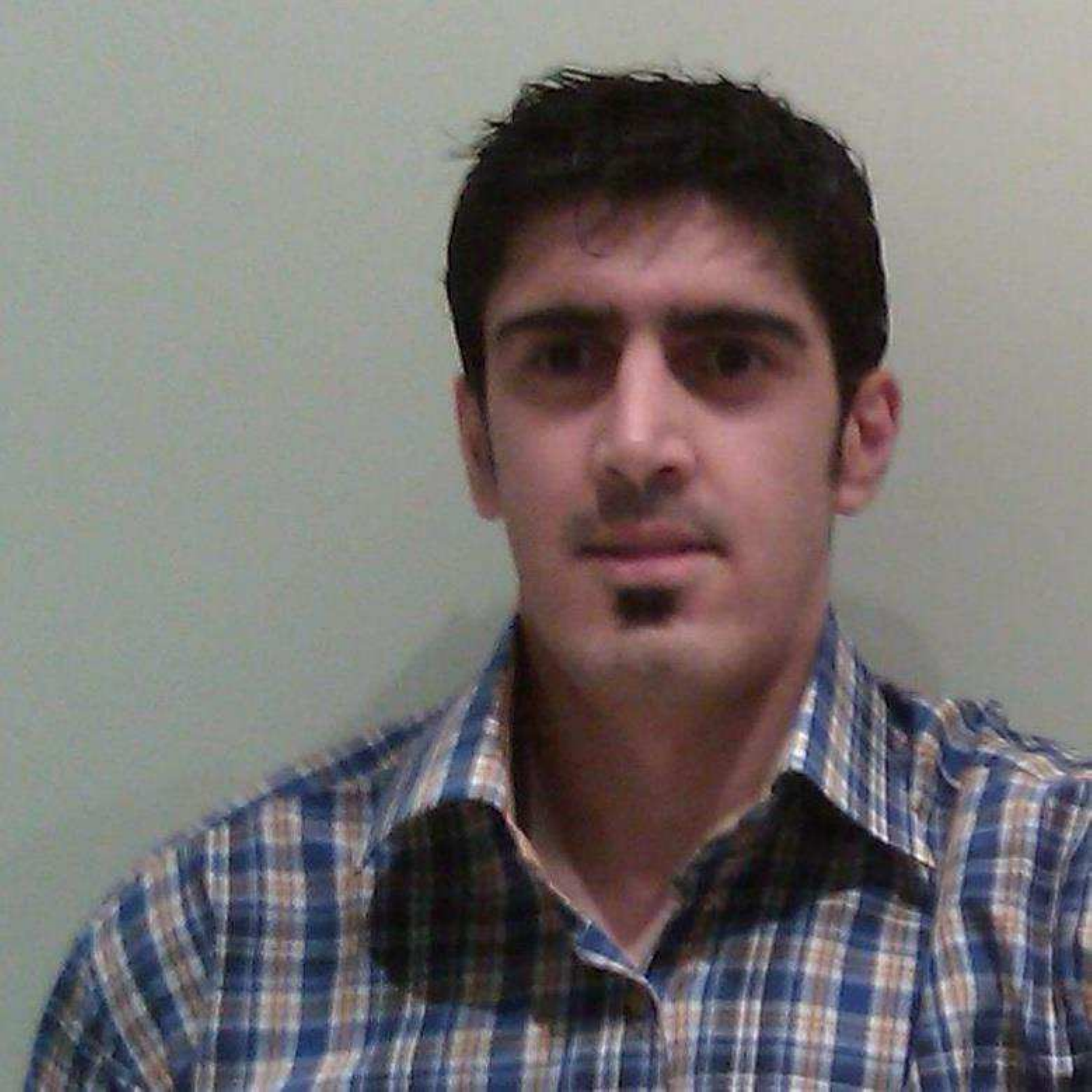}}]{Shahrouz Khalili}
received the B.Sc. and M.Sc. degrees from Shiraz University, Shiraz, Iran in 2009 and 2012, respectively, and his Ph.D. in electrical and computer engineering from New Jersey Institute of Technology (NJIT), NJ in 2016. He was awarded the best researcher of the year from Telecommunication Company of Iran in 2012 and the winner of the 2016 Hashimoto Prize for best doctoral dissertation in Electrical and Computer Engineering Department. His research interests include cloud computing, C-RAN, machine learning and signal processing.
\end{IEEEbiography}
\begin{IEEEbiography}[{\includegraphics[width=1in,height=1.25in,clip,keepaspectratio]{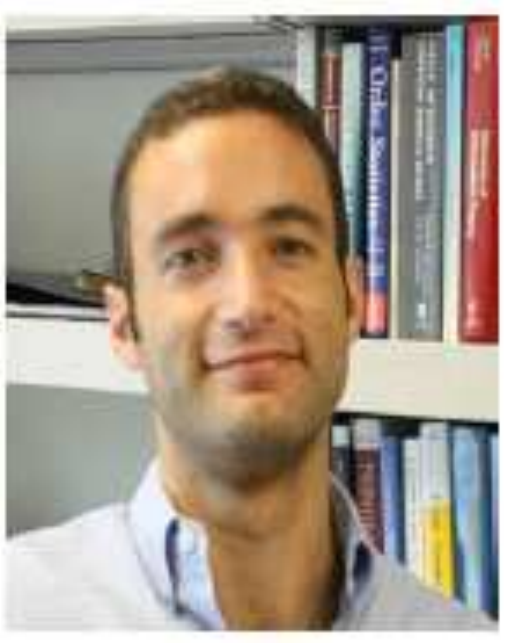}}]{Osvaldo Simeone}
received the M.Sc. degree (with honors) and the Ph.D. degree in information engineering from Politecnico di Milano, Milan, Italy, in 2001 and 2005, respectively. He is currently with the Center for Wireless Communications and Signal Processing Research (CWCSPR), New Jersey Institute of Technology (NJIT), Newark, where he is an Associate Professor. His research interests concern wireless communications, information theory, optimization and machine learning. Dr. Simeone is a co-recipient of the 2015 IEEE Communication Society Best Tutorial Paper Award and of the Best Paper Awards at IEEE SPAWC 2007 and IEEE WRECOM 2007. He currently serves as an Editor for IEEE Transactions on Information Theory. He is a Fellow of the IEEE.
\end{IEEEbiography}

\end{document}